# Visualization in virtual reality: a systematic review

**Elif Hilal Korkut and Elif Surer**

**Abstract** Rapidly growing virtual reality (VR) technologies and techniques have gained importance over the past few years, and academics and practitioners have been searching for efficient visualizations in VR. To date, emphasis has been on the employment of game technologies. Despite the growing interest and discussion, visualization studies have lacked a common baseline in the transition period of 2D visualizations to immersive ones. To this end, the presented study aims to provide a systematic literature review that explains the state-of-the-art research and future trends on visualization in virtual reality. The research framework is grounded in empirical and theoretical works of visualization. We characterize the reviewed literature based on three dimensions: (a) Connection with visualization background and theory, (b) Evaluation and design considerations for virtual reality visualization, and (c) Empirical studies. The results from this systematic review suggest that: (1) There are only a few studies that focus on creating standard guidelines for virtual reality, and each study individually provides a framework or employs previous studies on traditional 2D visualizations; (2) With the myriad of advantages provided for visualization and virtual reality, most of the studies prefer to use game engines; (3) Although game engines are extensively used, they are not convenient for critical scientific studies; and (4) 3D versions of traditional statistical visualization techniques, such as bar plots and scatter plots, are still commonly used in the data visualization context. This systematic review attempts to add to the literature a clear picture of the emerging contexts, different elements, and their interdependencies.

**Keywords** virtual reality · visualization · game technologies · systematic review

## 1 Introduction

The word “visualization” has been an overloaded term even before being established as a scientific field and has a prolonged usage with different meanings in different contexts. Since the visualization structures and types that can be presented in immersive environments are very diverse, immersive visualization is placed in the convergence of different research areas. In immersive environments, data can be presented with 3D models, 3D graphs and plots, simulations, and multiple 2D representations. The data source can be statistics, medicine, computer sciences, heritage, and many others. Its scope includes both technology-related areas, such as multisensory interfaces, interaction, navigation, collaborative aspects, rendering techniques, and domain-specific subjects.

Without a definitive starting event, the history of visualization includes many discussions collected around themes of design, purpose, or intent. Geometric diagrams, astronomical tables, and navigational graphics are

Elif Hilal Korkut
E-mail: elif.korkut@metu.edu.tr

Elif Surer
E-mail: elifs@metu.edu.tr

Graduate School of Informatics, Department of Modeling and Simulation, Middle East Technical University, Ankara, 06800, Turkey

considered the first visualization attempts, and prominent subjects of the field differ according to the era's problems and fields of interest. The increase in practical applications in the 17th century was closely related to interest in physical measurements, which led to more line graphs, astronomical graphics, and maps. For example, the first known weather map, a theoretical curve relating barometric pressure to altitude by Edmund Halley, and the plot of "life expectancy vs. age" by Christian Huygens were produced in that era (Chen et al., 2014). In addition to the increase in practical applications, with the collection of social data, demographic and economic visualizations were produced within the methods of 'political arithmetic.' The 18th century brought new domains and graphic forms, such as abstract graphs and thematic mapping. Joseph Priestley produced a more convenient timeline (1765) and a detailed history chart (1769). With the creative combinations of the fundamental forms, first-line graph and bar chart (1786), pie chart and circle graph (1801) were invented by William Playfair that are graphical representations still commonly used today (Friendly, 2007).

Most of the data representations used today took their form in the 19th century with the developments in statistical graphs. With the recognition of graphical representation by official and scientific spheres, graphical analyses were used in scientific publications and state planning. Together with the other innovative works of Charles Joseph Minard, his famous visual storytelling, the fate of the armies of Napoleon and Hannibal, are examples of the social and political uses of graphics which later gained the appreciation of most of the important names in the field (Rendgen, 2018). Among others, Nightingale's coxcomb plot, Jon Snow's Cholera map to enhance public health, statistical graphics and weather patterns of Francis Galton, and works of Karl Pearson are also typical examples of historic visualizations. After a fertile period, the early 1900s were defined as the modern dark ages of visualization. Analyses over time in the relational database of Milestone Project show a steady rise in the 19th century followed by a decline of the 20th century and until 1945, and continue with a steep rise to today (Friendly et al., 2017).

The insufficiency of traditional 2D representations leads the visualization community to search for more effective solutions. Recently, the interest in virtual reality (VR) technology and the contribution of interdisciplinary fields created new possibilities for application and implementation. VR is an immersive experience in an artificial environment. Throughout time, different methods and setups have been suggested for VR. The simplest version of VR is Desktop VR, a monitor display. Fish Tank VR includes both monitors and special glasses for stereoscopic viewing and uses the keyboard as the primary source of input. The Cave Automatic Virtual Environment (CAVE) is a surround-screen display technology consisting of room-scale projection surfaces to facilitate immersive virtual reality designed for exploration and interaction. The used projection technique allows users to see all directions. Immersive systems are mostly used with the help of Head-Mounted Displays (HMDs) today. HMDs are stereoscopic devices that display two images in front of the eyes to create a sense of depth. Depending on the technology used, interaction techniques can vary. Techniques can include head tracking, eye tracking, and motion tracking. Head-mounted devices present an opportunity for new data exploration and interaction methods.

In immersive environments, visual signals indicate an existence of a body movement while there is no actual movement, and as a result of this sensory conflict, cyber-sickness occurs. Different hardware has different frequency requirements to induce cyber-sickness. Degrees of Freedom (DoF) is a term used to describe the moving capabilities of an object. While basic HMDs provide 3 DoF for moving along the x, y, z-axis, more advanced devices offer 6 DoF, including the translational movement in physical space, surge, heave and sway. There are different interaction modes for VR. Users can have only a passive role or, most commonly, move with a pre-defined trajectory. Exploratory VR allows users to locomote themselves. In the interaction mode, users can explore the environment and interact to manipulate the environment, which is the most common interaction mode for immersive visualizations. Due to limitations of physical spaces, HMDs provide seated configurations, allowing users to move with controllers and room-scale VR. VR recreates a spatial environment and builds three-dimensional spatial awareness via visual cues and sounds.

Immersive environments generally refer to certain terms, such as presence, immersion, and embodiment. Sense of embodiment depends on the spatial components provided for the user, such as awareness of location and virtual body. Presence is related to being in the virtual environment, and immersion can be considered as the result of this presence. The combination of immersion, presence, and embodiment contributes to the user's experience and determines its quality. Therefore, they are widely used to evaluate and develop VR experiences. Most of the studies employ questionnaires to measure presence and immersion.

Other than the building setups, creating large-scale VR environments was problematic due to the absence of software tools. Recently, game engines Unity and Unreal

Engine have been widely used to build VR environments. Rapid production offered by game engines has allowed many areas to build immersive visualizations. Condense information extracted from data needs to be presented in a visual form. This form can be animated, static, or interactive. The definition of data type and selection of visualization and interaction techniques is crucial to create efficient and accurate visualizations. The selection of appropriate presentation techniques depends mainly on the user. Therefore, visualization techniques depend on perception and cognitive theories to convey the data efficiently.

As an interactive communication method, visualizations are expected to provide certain features and tasks, such as presentation of the data and confirmatory and exploratory analysis. Data visualization and exploratory data analysis have gained tremendous importance in recent years due to the increasing amount of data. Extracting information from high dimensional and large volume data required visualization domain to employ different automation techniques such as machine learning algorithms. Recent advances in immersive technologies and computational power present new possibilities for data exploration methods that aim to provide interaction with high-dimensional data to gain fundamental insights. With the increasing capabilities of the hardware and software and the necessities of the time, VR devices have become more useful and affordable. Immersive technologies change data experiences and decision-making press processes. It allows users to analyze the complex and dynamic dataset and change their passive roles to active ones.

Visualization subfields, besides common problems of VR, have domain-specific design problems. The creation of information visualizations includes a decision-making process regarding abstraction methods, visual encoding, and design principles. Scientific visualizations need to deal with the problems of scalability, accuracy, and precision. Visual Analytics (VA) is concerned about activities that can be performed through visualizations, such as decision making and reasoning. Immersive Analytics (IA) focuses on using display and interface technologies to support better analytical reasoning and decision-making processes.

Visualization and interaction opportunities provided new ways to express ideas and propose new interaction methods for a wide range of research domains and disciplines (Figure 1). With recent technological advances, the invention of several libraries, tools, and devices, VR facilitates the manipulation and analysis of data by using the advantages of 3D environments. Combining VR and haptic or kinaesthetic interfaces enable various interaction techniques and maximize efficiency. The generation of immersive visualizations has further improved various domains in terms of practicability, education, and cost-effectiveness. Digital city technology allows users to create more sustainable and effective solutions for urban environments. Providing reusable and safe environments for experiments and education, virtual environments provide training in diverse areas.

VR and 3D immersive environments are perception-related technologies that need to have their visual language. Therefore, it is necessary to continue building theoretical approaches. Ensemble of studies has the power to lead a groundwork for visualizations. The bidirectional contribution and the influence between games, video games, and VR have propelled the extension of VR into scientific, artistic or informational, and educational domains. While most VR studies embrace a ludic approach, VR has created an ocean of possibilities that contain new mechanics, narratives, and interactions for the game industry. Zyda (2005) strongly advises VR researchers to study games to improve their design and stay up-to-date.

While a wide range of areas adopts this technology, it is crucial to think critically about the solution for challenges, visualization specifications, and design guidelines. Most existing surveys on visualization in virtual reality focus on a specific domain or specific visualization structure. For example, Zimmermann (2008) focuses on the automotive industry and design aspects. Seth et al. (2011) explained assembly methods for prototyping, Radianti et al. (2017) focused on higher education, Wang et al. (2018) surveyed construction engineering from education and training perspectives, El Jamiy and Marsh (2019) inspected depth estimation, Caserman et al. (2019) provided studies and analysis on full-body motion reconstruction, and Ferdani et al. (2019) analyzed the studies on archaeology. However, considering the specific requirements of immersive technologies, visualization techniques that are used in different domains are closely interrelated and have the potential to create a mutual relationship to solve problems of immersive technologies. Previously, while data collection was the problem, knowledge extraction and presentation became ubiquitous due to massive data. To fully achieve the visual representation objectives, the construction of visualization methods depends on various areas, from psychology to machine learning. Therefore, we survey techniques from a broader perspective to extract the relations, similarities, and shared problems in visualization in VR. We think that this approach can help developers find solutions in other domains and give a direction to more concrete guidelines. This paper aims to present an overview of existing literature and discuss the common problems and methods used in different domains to provide a basis for the immersive visualization domain to construct a comprehensive and consistent structure.

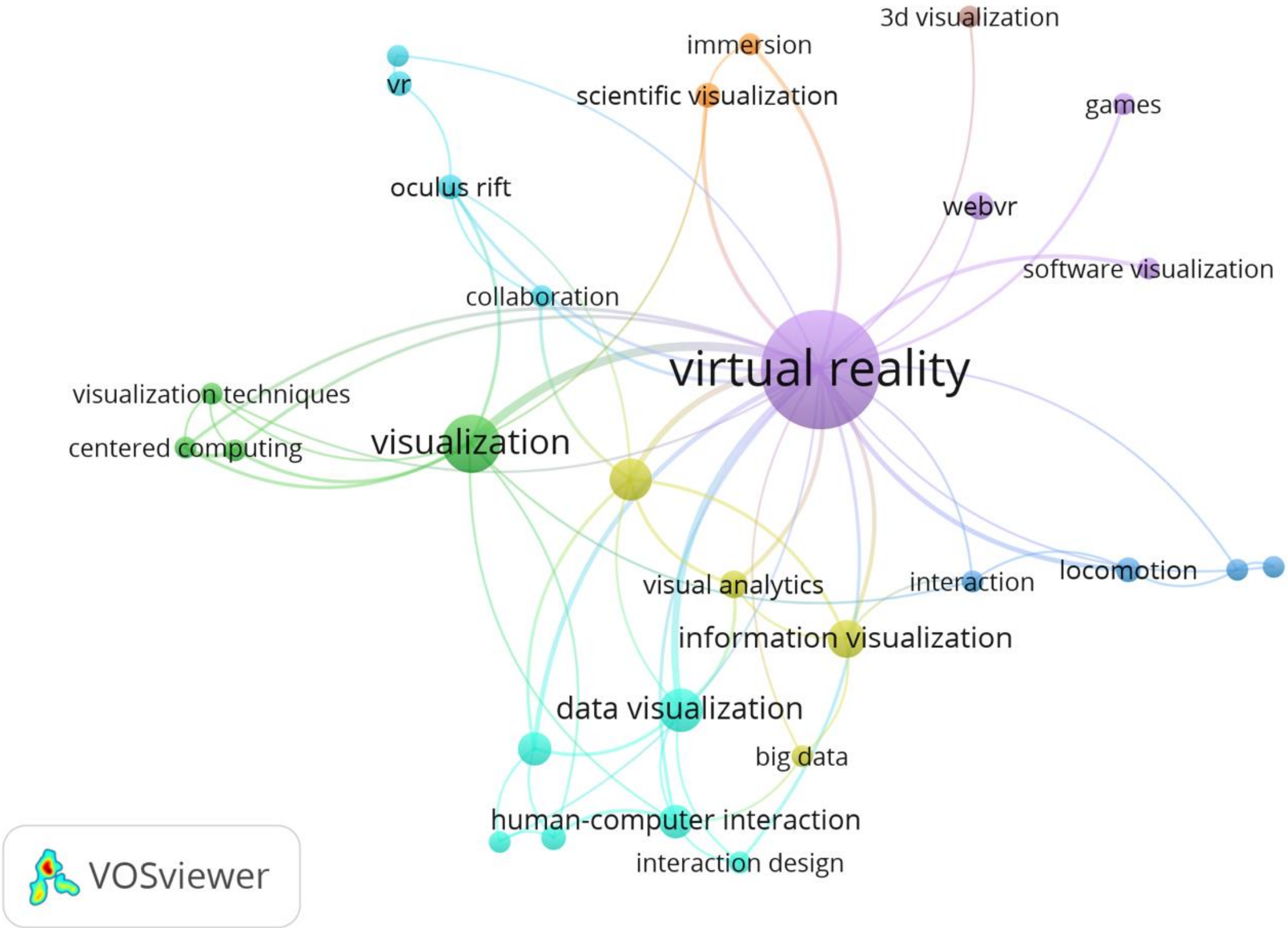

**Figure 1** Co-occurrence Keyword Network Produced with VOSviewer

The remainder of this study is structured as follows. While Section II presents background concepts, Section III briefly describes the methodology, Section IV summarizes the results, and Section V presents the conclusion.

## 2 Background concepts: Visual Analytics and Immersive Analytics

The emerging field of visual analytics is explained as “the science of analytical reasoning facilitated by interactive visual interfaces” (Keim et al., 2008). With the massive volumes of information waiting for human judgment, it is seen as a critical technology to handle big data (Mehrotra et al., 2017). The urgent need to analyze complex data leads to the integrated work of the user and the computer. For example, visual analytics systems are actively studied in medicine to provide better healthcare. As part of the Electronic Health Records (EHRs), clinical decision support systems (CDSS) (Moon and Galea, 2016), interpretable machine learning for recurrent neural networks (Kwon et al., 2018), supporting comparative studies of patient records (Guo et al., 2020), and applications such as OutFlow (Wongsuphasawat and Gotz, 2011), CarePre (Jin et al., 2020) and EventAction (Du et al., 2016) are used. To provide a better understanding, convolutional neural networks (CNNs) were also studied with visual analytics systems (Liu et al., 2016; Jacovi et al., 2018; Chawla et al., 2020). While visual analytics is more concerned with getting insights, detecting interesting patterns, and gaining a deep understanding from visually represented data, a new term has emerged with the development of 3D-based data exploration tools. Chandler et al. (2015) define the phrase 'Immersive Analytics' as “the use of engaging embodied analysis tools to support data understanding and decision making.” Combining data visualization and visual analytics with technological developments, immersive analytics aims to remove the obstacles between humans and data for making all processes available for everyone.

The increase in the use of immersive and spatially oriented technologies, including virtual, augmented, and mixed reality (VR/AR/MR) devices, creates a desire to explore the potential for complex data sets within a collaborative and interactive environment. With the increased accessibility to the technology, those devices have started to be used by non-specialists who create a need for the field of Human-Computer Interaction (HCI), where psychology and computer science blend together. Although Ivan Sutherland demonstrated virtual reality and augmented reality prototypes in the late 1960s, the practical, interactive virtual reality systems had to wait until the 1990s (Dwyer et al., 2018). In 1992, Cruz-Neira et al. (1992) reported the CAVE system as an early example of an immersive virtual reality system. Enthusiasm for interaction with virtual content followed by a series of events. Thereby, as a unifying term, immersive analytics had a chance to merge the areas of Immersive Information Visualization, Visual Analytics, virtual and augmented reality, and Natural User Interfaces successfully.

Recent research includes the development of new techniques for visualizations, interaction and collaboration, evaluation of perception and systems, proposals for frameworks and tools definition, and categorization of challenges. For instance, by extending the work Brehmer and Munzner's (Brehmer and Munzner, 2013) What-Why-How framework for immersive analytics, Marriott et al. (2018) propose to use Where-What-Who-Why-How questions as a basis.

## 3 Materials and Methods

This study offers a systematic literature review of the studies related to immersive visualizations in virtual reality. The research questions of this study are:

RQ1: What are the most preferred visualization types and structures for VR visualization?

RQ2: What methodologies/theories are being used to research VR visualization?

RQ3: What are the research gaps in VR visualization?

RQ4: What are the existing approaches and techniques?

RQ5: Which software and hardware have been preferred for different types of visualizations?

We have performed a systematic literature review following the guidelines proposed by Kitchenham and Charters (2007) to answer the abovementioned research questions. To find relevant research that has been published since 2015, we selected seven primary academic databases: ACM Digital Library, IEEE Xplore, SpringerLink, Science Direct, Google Scholar, Elsevier, and Web of Science. We broke the query down into the major research fields to perform the search. The primary terms were 'virtual reality ' and 'visualization.' After considering alternative spellings, search terminology included related areas combined with virtual reality. The resulting search strings were: visualization/visualisation AND virtual reality, virtual reality AND ("data visualization" OR "information visualization" OR "information visualization"), immersive visualization, immersive AND visual analytics, virtual reality AND game. We searched for the title, abstract, and keywords with this query. The search process was carried out between August 2019 and February 2022, and the timeframe for publications was between 2015 and 2022 (Figure 2).

### 3.1 Inclusion and Exclusion Criteria

After performing the searches, due to the importance of the selection phase, each candidate study was subjected to a set of stages composed of inclusion and exclusion criteria. Publications were selected from 2015 onward, and the systematic review included journal articles, conference proceedings, in-progress research, and scientific magazines. Online presentations were excluded. Initially, we assess the relevance of literature according to the titles and the abstracts, looking for papers that described either virtual reality visualization or its indicators. Papers that do not include ques about visualization elements or techniques were excluded. Secondly, we retrieved each study, read it entirely, and critically appraised it based on similar criteria as stated above. Then, validated studies were grouped according to association levels with visualization subfields. Finally, we removed duplicate papers or preliminary versions of works already being analyzed. A total of 1067 papers were obtained by running the query in the databases. Out of these 1067 papers, 474 papers passed the first stage. After removing non-relevant papers, 220 papers contained related studies, and 94 were found as primary sources. Additional to those papers, studies published before 2015 were included, which present a theoretical proposal to explain the motivation behind the implementation studies.

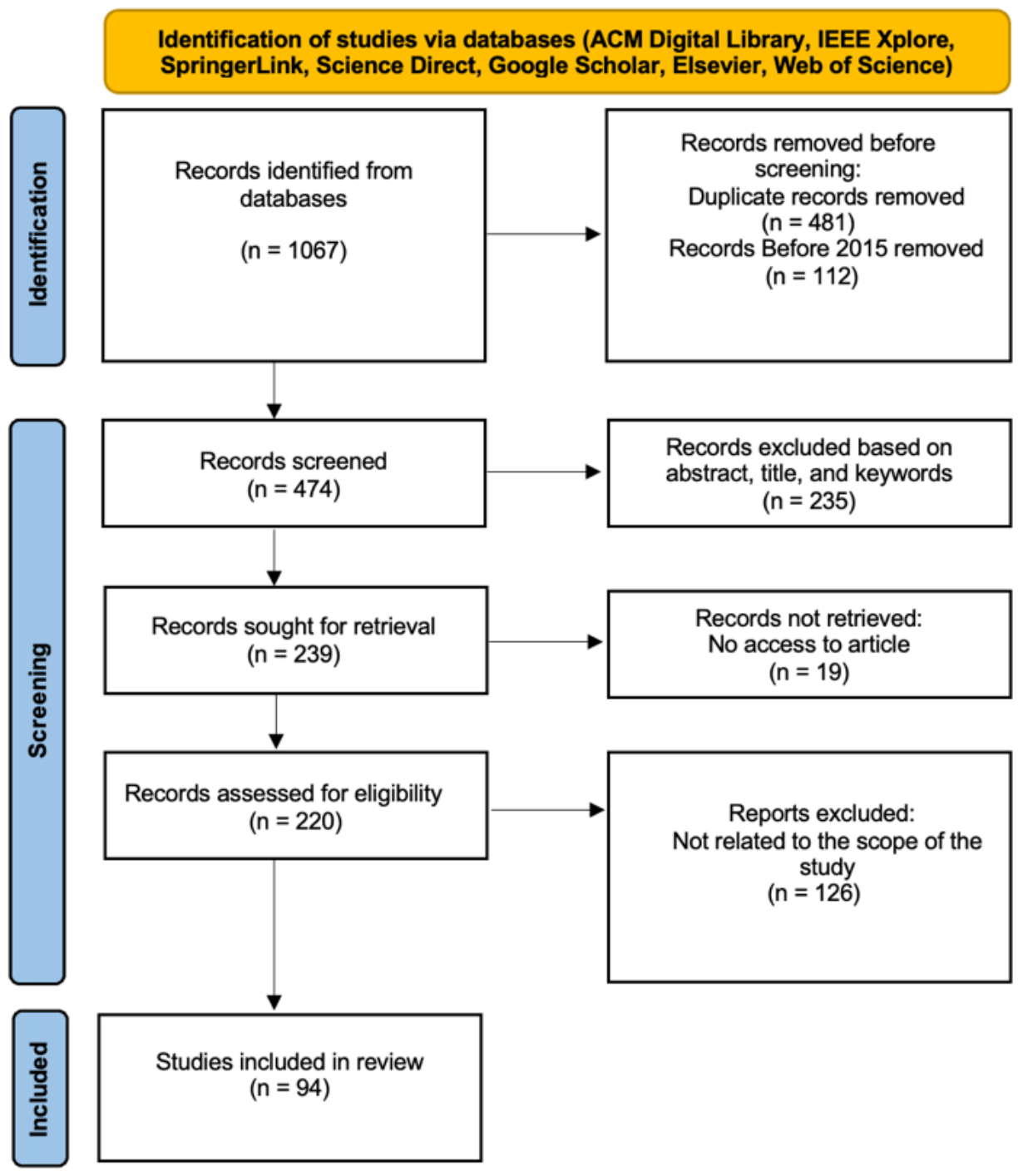

**Figure 2** PRISMA Flow Diagram for systematic records selection

**3.2 Types of Contributions and Division**

The review distinguished between two types of contributions. The first type includes the papers that contributed a theoretical proposal or a framework. The second type involves most papers that describe the implementation of visualization. Papers that contributed a combination of a theoretical framework and a subsequent implementation were included in different sections of the results. Sections of this survey were constructed according to condensed areas or extensively used terms (Figure 3). Although in some cases category of the study is clear or defined by the authors, due to interwoven feature of visualization, a study can belong to more than one section. For such cases, the category of the study was clarified according to definitions of the terms given in the studies of Rhyne et al. (2003) and Kosara et al. (2003). After an extensive literature search and review, the resulting research papers were grouped around ten main categories

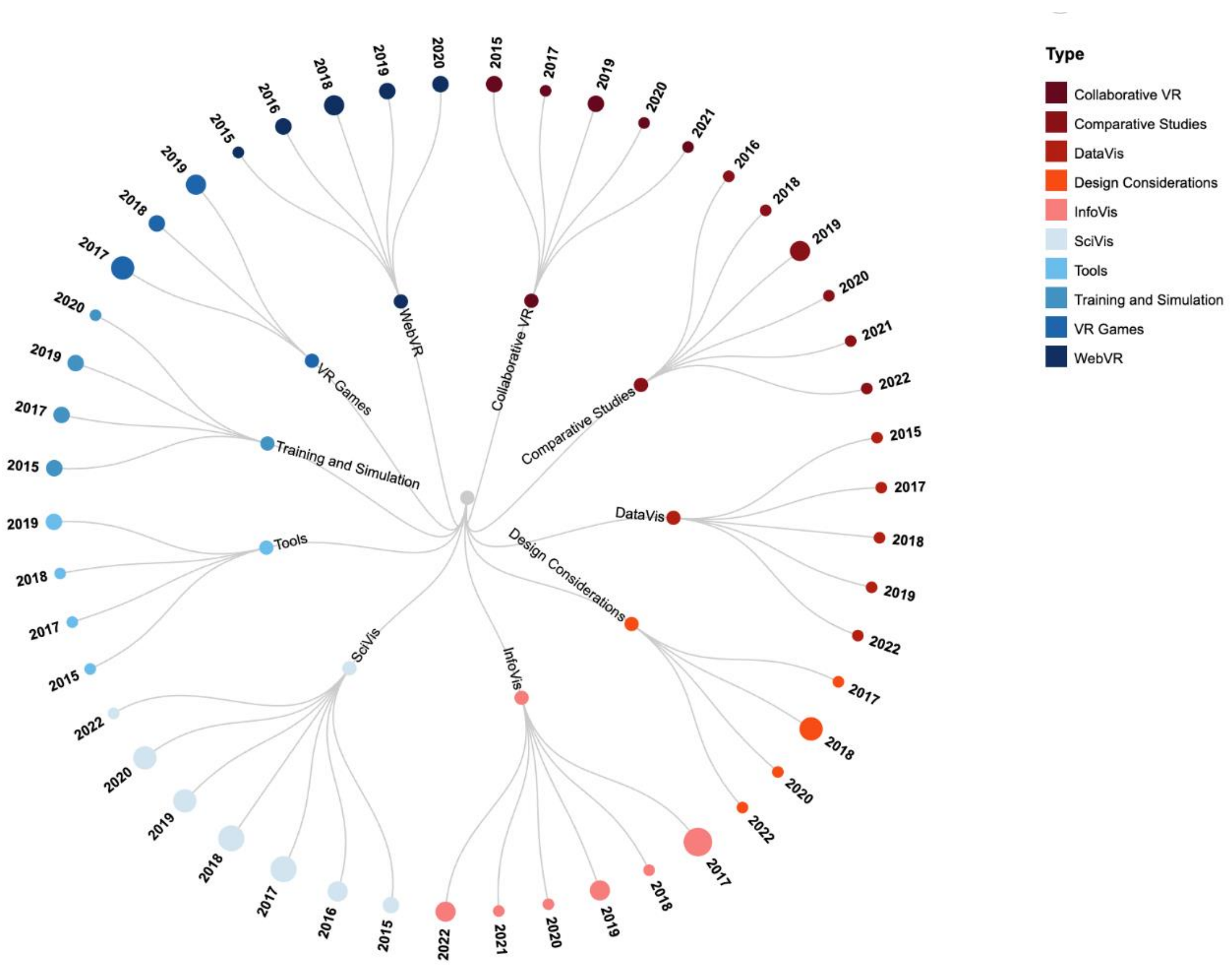

**Figure 3** Distribution of studies Produced with RawGraphs

according to their contribution, domain, and visualization category.

## 4 Results and Analysis

### 4.1 Tools, Toolkits, and Frameworks

Visualization tools can be generally standalone, web-based presentations, web-based development mainly consisting of software libraries (APIs), or programming language modules (e.g., Python or Java module). They can also be categorized in terms of software, visualization structure, operating system, license, scalability, extendibility, or latest release date. According to the criteria above, Caldarola and Rinaldi (2017) reported 36 software tools grouped into four subsections; scientific visualization, data visualization, information visualization, and business intelligence tools. Database-related and GUI-based applications provide "direct manipulation principle" such as Microsoft Excel, Amazon Quicksight, and Microsoft Power BI. Although they are widely used, since they are out of the scope of this article, further detail will not be given. Visualization construction tools are generally criticized for preventing creativity due to fixed properties; however, they are preferred since they provide easy-to-use environments without requiring programming.

Although the visualization libraries reduce the complexity, they still require experience. In addition to those, there are development platforms and existing cross-platform tools whose scope involves multiple areas. Need for easy-to-use and flexible graphical systems to support visual thinking paved the way for further developments. Beginning with Bertin's Semiology of Graphics (Bertin, 1983), formalization of the graphing techniques has started, which later transformed into structural theories of graphics to establish a bond between computer graphics and information visualization theories. Cleveland and McGill (1984) experimented with retinal variables (position, color, and size). More recently, the ideas and theories of Wilkinson (2012) provide a basis for visualization interfaces Lyra (Satyanarayan and Heer, 2014) and VegaLite (Satyanarayan et al., 2017), and grammar-based systems such as Polaris (Stolte et al., 2002), which extends the pivot table interface. Visualization production tools such as Lyra and iVisDesigner (Ren et al., 2014) enable the creation of a variety of customized graphic visualization based on conceptual modularity without writing any code. Unfortunately, they support only a small set of visual forms and parameters that limit users.

Drawing from Wilkinson's grammar with the formalization of the grammar of graphics, many visualization grammars, toolkits, and frameworks have been implemented. Those declarative languages are generally grouped into low-level grammars and high-level grammars. Low-level grammars such as D3 (Bostock et al., 2011), Vega (Satyanarayan et al., 2015), Protovis (Bostock and Heer, 2009), are expressive grammars to help designers to create explanatory and highly customized graphics with fine-grained control for data visualization where all mapping elements need to be specified. More recently, D3 became very popular, especially for web development. Protovis is an embedded domain-specific language implemented in JavaScript, and defining graphical marks such as bars, lines, and labels helps users specify data bindings to visual properties. Vega is similar to Protovis and D3, but it provides transformation on scales and layout with support modules, and interactive connection between input data and properties of the marks allow users to share and reuse the product. On the other hand, high-level declarative grammars such as Vega-Lite, and ECharts (Li et al., 2018) are better for exploratory visualization, and by encapsulating details and properties, they focus on the rapid production of visualizations. A declarative statistical visualization library for Python was developed named Altair (Satyanarayan et al., 2017).

Creating visualizations using APIs requires background knowledge, and it is a grueling process. Therefore, frameworks have been created for rapid and better abstractions. After introducing InfoVis, toolkits that provide a collection of visualization tools similar to Java-based visualization libraries were developed, such as Prefuse (Heer et al., 2005). In addition to provided operators and abstractions of libraries, Prefuse and Flare (Gal et al., 2014) allow users to define new ones and with fine-grained monolithic units to provide the customization. GPU-powered visualizations have been widely used in scientific visualization, and their use in information visualization increased in recent years due to improvements in rendering performance. For example, Stardust (Ren et al., 2017) utilizes those improvements. It does not provide a new visualization grammar, but it is a complementary work for previous tools with more user-friendly building blocks, and it enables the creation of both 2D and 3D visualizations. Pointing out the gap between artists and the expert coders, a programmable integrated development environment (IDE) named VisComposer (Mei et al., 2018) has been developed, which uses tree-based visual structures similar to D3. VisAct (Wu et al., 2020) is another interactive visualization system that provides a high-level grammar for semantic actions and guides the users by including a wizard panel and a wide range of visual forms.

The effort to construct interactive toolkits or systems for information visualization had limited itself to 2D representations that are more traditional. Therefore, with already suitable 3D environments for immersive environments, scientific visualization has led to the development of virtual reality systems. A widely used framework targeted at scientific visualization applications is the Visualization Toolkit (VTK) (Hanwell et al., 2015), an extensive library for displaying and interacting with data. Using VTK in the VR environment became possible with the development of OpenVR. This API supports SteamVR developed by Valve. Thus, the framework became compatible with Oculus Rift and the HTC Vive.

Researchers have recently paid more attention to exploring immersive environments for non-spatial data. Although designed for gaming applications, the Unity game engine has become a standard platform to develop immersive environments. Both IATK (Cordeil et al., 2019) and DXR (Sicat et al., 2019) toolkits were developed for building immersive data visualizations based on the Unity game engine. DXR is a toolkit that uses a declarative framework inspired by Vega-Lite and provides interaction and extendable visualizations with additional classes and applications that can be exported to various platforms, including mixed reality (MR) on Microsoft HoloLens, and VR headsets. On the other hand, the API of IATK is similar to D3, and using a grammar of graphics allows easy construction of visualizations. Emerging from the previous applications like ImAxes (Immersive Axes) (Cordeil et al., 2017a) and FiberClay (Hurter et al., 2018), IATK allows users to create visualizations by three-dimensional axes, but it does not provide collaboration. Fiberclay, being a flagship example, was evaluated with air traffic controllers, and it displays large-scale spatial trajectory data in 3D, where it provides selections of 3D beams for constructing queries. ImAxes is an open-source information visualization tool that implements scatterplots, histograms, and parallel coordinates that are explorable based on manipulating reconfigurable axes using natural interactions.

### 4.2 Data Visualization

Data visualization represents data or information in a graphical format that enables the audience to identify patterns, pull insights, grasp the true meaning of information, and communicate more quickly and efficiently (Aparicio and Costa, 2015). While diverse areas benefit from the graphic representation of data, data visualization also feeds on several disciplines. Transformation of the data into compact and understandable information in pictorial format became possible with the contribution of psychology, computer sciences, statistics, graphic design, and many other disciplines. Flourishing with the knowledge from multiple backgrounds, adaptability and scalability of the data visualizations have increased. For example, continuously accumulated data in various domains eliminates the traditional methods, which are currently insufficient for large batches of data. Different methods, such as machine learning, can be applied to conduct analyses and create more efficient visualizations with varying attributes. Existing 2D methods of data visualization can contain only a small number of correlations between a few metrics. Thus, to perform analyses on high-dimensional data, many individual charts are required for comprehensive presentation, eventually preventing comprehending correlations and patterns.

Direct conversion to 3D does not offer enough clarity since problems of 3D such as occlusion and perspective distortion may lead to wrong interpretation in analytical use cases. Although 3D graphics can be effective, they can be considered unnecessary according to data and visualization structure. To enhance the data visualization experience, it is necessary to have additional techniques to display information in greater depth. For example, according to a survey (Fonnet and Prie, 2021), position and visual channels such as textures, colors, and shapes are commonly used to encode multidimensional data.

Interactivity is one of those aspects that enhance 3D environments. Virtual reality changes how we interact with and interpret data, and visualizations should support several activities. VR should enable an exploratory analysis to discover the input data and its features, tendencies, and relations. To help the user to reject or accept the constructed hypothesis, it must offer confirmatory analysis. The presentation of data should be given in a structured manner to reveal the hidden features which cannot be presented via other mediums or platforms.

Sun et al. (2019) provided dynamic visualization of the time series along with geographical attributes and made visualizations available to observe the relation between accumulation, wind direction, time, and location. They used an aggregation table, calendar view, day bar graph, and line plots to visualize data coming from air sampling sensors and meteorological data. In the study of Okada et al. (2018), the visualization system generated for spatio-temporal data is composed of two layers. The first layer presents a spatial model with an adjustable scale according to worldview and minimap options. The second layer represents the frequency with cubes with different colors and transparency. The combination of multiple visualization techniques in single VR visualization improves the information flow and creates more engaging experiences.

As a communication medium, another critical element of visualization is interpretation. Design and interpretation choices of visualization can alter the users' ability to comprehend the data or lead to misunderstandings. Therefore, a good visualization should protect the balance between aesthetics and functionality. Graph layout algorithms and clustering algorithms are extensively used for complex network visualizations. Clustered data needs to be converted into comprehensible visualizations. For example, Drogemuller et al. (2017) preferred spheres for entities, lines for relationships, and circles for cluster nodes to construct a network visualization using a spring embedder layout. Clustering algorithms help users detect patterns quickly and assist them in inspecting high-dimensional datasets. However, utilized algorithms may have poor performance because of dimensionality or noisy data. Therefore, while presenting a refined version of the Immersive Parallel Coordinates Plots (IPCP) system, Bobek et al. (2022) prefer to use a wide range of clustering algorithms for multidimensional datasets. Also, they show the importance of feature selection by testing multiple feature selection methods.

Unlike the 2D data visualizations, where the data is always presented from one perspective, VR exhibits potential uses for switching between perspectives. This creates different embodied cognition cases for users to interpret data in new ways. The ability to change perspectives creates more immersive experiences and precise insights. The focal point changes according to the user's movement in a virtual reality environment. Therefore, according to the user's perspective, the perceived distance of contents varies. The layout of the presented data can provide equivalent perception and ensure distribution in spherical space; Kwon et al. (2015) employ a space-filling curve layout together with spherical edge bundling. Different strategies, algorithms, alternative ideas, and presentations, the combination of those techniques enhanced the visualization. The creation of elaborative interpretations of complex data allows users to gain a deep understanding by seeing data in different ways.

## 4.3 Information Visualization

Exploring the design space of spatial mappings for abstract data became a key theme of a new sub-field of visualization called Information Visualization (InfoVis). Information visualization was built upon graphic designers, statisticians, human-computer-interaction (HCI) researchers, and many others. The interdisciplinary field has been exploring the effective use of computer graphics for abstract data visualization and its interactive exploration (Figure 4).

### *4.3.1 Art, Heritage, and Architecture*

Thanks to the gaming industry, with the recent development of low-cost devices which can provide powerful VR and augmented reality (AR) applications, cultural heritage institutions have started the digitalization era. 3D representations of cultural heritage artifacts and structures have affected many areas, such as virtual cultural heritage tourism and research on urban history. Development of the technology which can capture and document the heritage sites offer new techniques that substitute the human interpretation, which requires more time and workload. After the first 3D documentation of archeological objects was realized by Leo Biek (Gettens, 1964), many artifacts started to be exported to digital environments. With sufficient documentation for transmitting the cultural heritage to future generations, which cannot be protected due to natural and artificial disasters, preservation in digital mediums has become a reliable method. Final products of the digitization process do not only serve for archaeological and architectural documentation, but they also provide educational opportunities, exhibitions, virtual tourism, experimental studies on space, and analyses on artifacts. On the other hand, preparation, presentation, and interaction of 3D digital contents require meticulous work. According to the complexity, scale, and location of the subject of 3D digital representation, different methodologies can be used, such as laser scanning and photogrammetry. Also, several techniques have been developed to reconstruct an artifact, such as sculptures and paintings. These techniques include image sequencing, volume-based methods, structured form motion algorithm (Sooai et al., 2017). The problem is that generated models are generally complex due to graphic requirements of details. Several geometric optimizations and compression methods have been developed to solve technical problems such as managing millions of polygons or processing time, while their main aim is not to comprise details and realism. For example, Fernandez-Palacios et al. (2017) offered a pipeline including many optimization techniques to create an immersive VR experience with digitally reconstructed heritage scenarios. Their work includes normal maps to transmit details to low-resolution models, unwrapping techniques to decrease the texture load, and the use of software tools to decrease the resolution for geometric optimization without reducing visible quality. Choromanski et al. (2019) established a VR system that utilizes terrestrial laser scanning, and photographic images belong to a baroque palace. They also tested various texture mapping algorithms to simplify the mesh geometry of models constructed with data gathered through different methods.

Another way of protecting and maintaining cultural heritage with digitization is virtual museums. Schweibenz (1998) defines virtual museums as a collection of relevant digital objects to disseminate objects and information without real place or space. The construction of virtual museums is challenging due to its requirements, such as user interaction, environment and content, and design of the experience. The museum can be designed in realistic form, duplicate an actual museum, or alternative methods can be used, but the final product should convey the intended information (Skamantzari, 2018). For example, besides the realistic models, the virtual museum design of Kersten et al. (2017) includes guided viewpoints for essential positions and detailed information menus. Monaco et al. (2022) created a customizable virtual museum where users can have a more active role in virtual exhibitions. Their construction process allows users to select data using Knowledge Graphs (KGs), personalize the museum by changing the layout, and select annotations. Knowledge is a set of entities, properties, and relations between them. Entities, relations, and properties can form a graph of nodes and edges, making graph structure a realistic representation of knowledge. Combining virtual exhibitions with the graph abstractions enhanced the interactivity and made the complex relations of knowledge understandable in a context that cannot be presented in traditional museums. They also reported that construction time is highly related to the lighting settings after users' selections since it requires heavy computation. Therefore, they offer alternative lighting options.

The assemblage of VR technology and geographic information system (GIS) generated a new information system called VRGIS. VRGIS can support spatial data query, processing, storage, and analysis functions. Combining multiple technologies, such as Internet of Things (IoT) technology, VR, and 3D geographic information system (3D GIS), provides new ways for producing sustainable urban environments. Compound use of visual analytics and GIS systems allows understanding important features, such as betweenness, closeness, centralities, and shortest paths in the urban design domain. However, the city scale and components generate graphs that are not workable to explore or understand. One way to solve this clutter problem is simplification and division of graphs. For example, Huang et al. (2016) generated a visual analytics system called TrajGraph to study and plan urban networks. They applied a graph partitioning algorithm to divide graphs into several chunks while conserving the necessary relations for objectives. The smart city concept is proposed to optimize urban systems and form sustainable and efficient environments. According to Lv et al. (2016) construction of a smart city is composed of information, digitization, and intelligence stages. Broucke and Deligiannis (2019) propose a VR platform on Brussels' smart city data, which shows a reduction in frustration level in exploratory experiences of participants. Dong et al. (2022) conducted a detailed analysis to understand virtual reality requirements for multilayered data of cities. In parallel to the results of their analysis, they constructed a digital city simulation model based on multiple components and subsystems, such as model editing and restructuring models (MERM) and scene creation and roaming system (SCCM). Their process starts with the data collection, and it convolves into different formats to create a consistent platform. Most smart city projects aim to improve urban life and create environments to support efficient and effortless interaction for urban dwellers. The smart city concept also can be helpful in urban construction. In the construction process, it is crucial to foresee practical issues. It is also essential to understand spatial order, functions that are applicable, technical requirements, and production process. Considering the parameters and scale of a city, 3D visualizations and simulations can orient the decision-making process.

To make art experiences more accessible, immersive, and engaging, digitally-based strategies such as AR, VR, and Web3D have been employed. Recent projects mainly focus on the recreation of an artifact in a virtual environment. Even though the results of many studies are promising, the process of implementing an artist's universe is not an easy task. For example, transmitting a 2D art form into VR experience requires developers to add the parts of the painting not included in the original work, and duplication requires every detail to be modeled in 3D. Raya et al. (2021) reconstructed two paintings for VR, where the paintings became available for users to experience kinesthetic textures with an introductory approach. According to a review paper (Zhang et al., 2020), VR technologies are employed in architectural domains in safety planning, design interpretation, collaboration, construction project management, education, planning, and human behavior and perception. In recent years, integrating Computer-Aided Design (CAD) and Building Information Modelling (BIM) tools with VR has been promoted to maintain efficient communication and design processes and avoid conflicts for the areas listed above. Architects and engineers use BIM for efficient design, management, construction, and operation stages. For example, on the effects of daylighting, Akin et al. (2020) developed an immersive design tool integrating BIM technology to improve visual perception and awareness during the design process. The 3D CAD models contain a large amount of information in 3D models, 2D drawings, and charts (Ivson et al., 2020). Munster et al. (2020) offer an automated pipeline to construct 4D city models from historical images to create browser-based VR applications for mobile, where the fourth dimension here is time. Utilizing a CNN architecture, they generated models of buildings from images based on their floor plans. Attempts to engraft different fields also led to a wide range of subfields, such as immersive urban analytics. For example, Chen et al., 2017 proposed a method to apply visual analytics in the urban context using the exploded views and principles explained by Li et al. (Li et al., 2008).

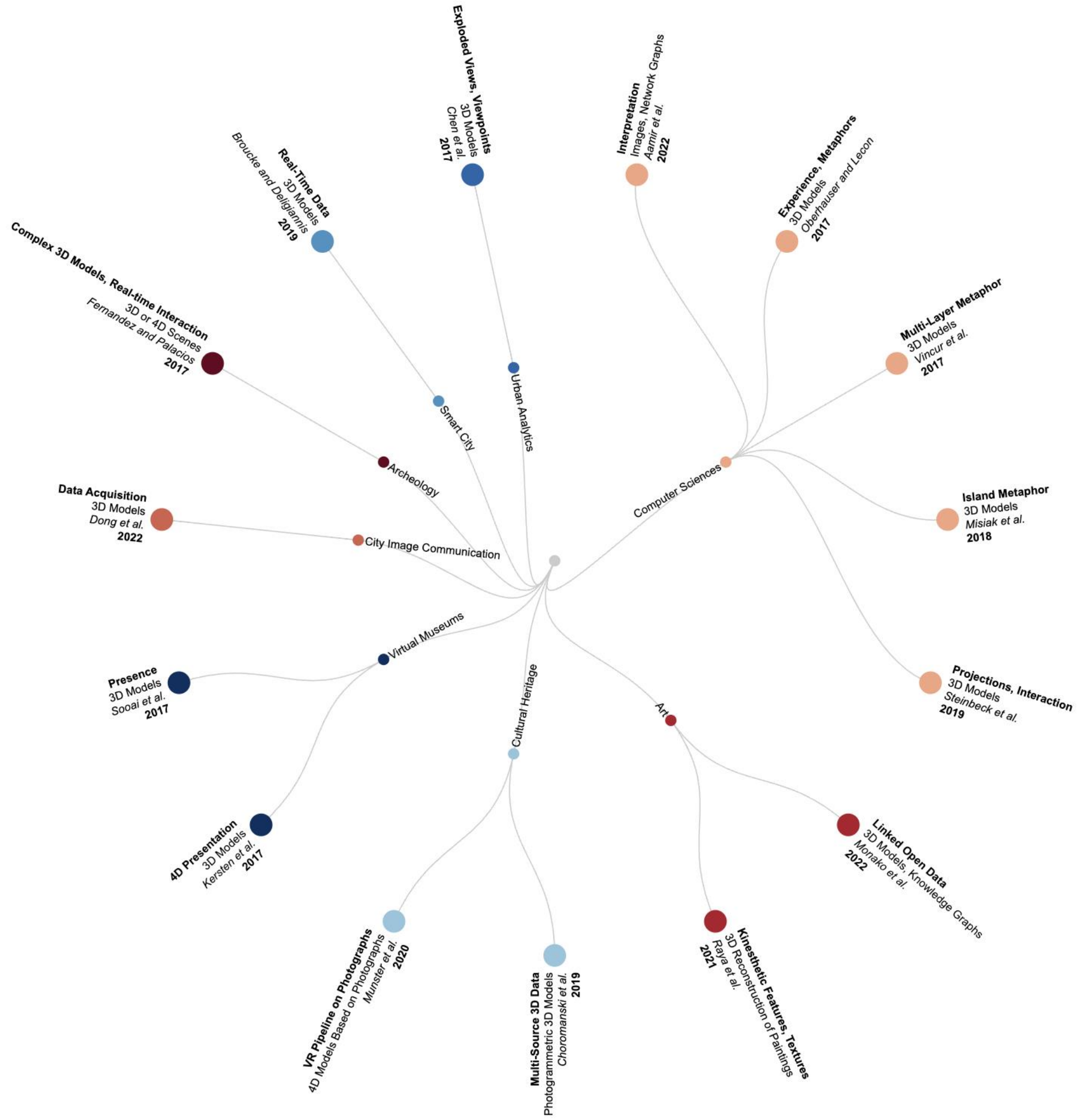

**Figure 4** Distribution of Information Visualization studies Produced with RawGraphs

### *4.3.2 Computer Sciences*

Different visualization techniques have been offered to better understand software architectures, various algorithms, and computer science concepts. Studies focus on interpreting complex structures to understand different features and concepts related to the field. For example, visualization techniques have been employed to better understand and explain artificial intelligence (AI). Explainable Artificial Intelligence (XAI) is a recently developed technology aiming to enhance the understanding of AI from the eyes of humans. As part of this study, Selvaraju et al. (2017) offered to use Gradient-weighted Class Activation Mapping (GradCAM), which is a visualization method that benefits from the gradient of the target and produces localization maps on Deep Reinforcement Learning (DRL) algorithms. With the analysis of their study on Atari Games that includes visualizations of input states and selected output action, the role of the CNN layer can be understood. Another study that focuses on visualizing neural networks is Caffe2Unity (Aamir et al., 2022). Combining the Caffe framework with the Unity game engine provides real-time interaction with the neural network on the image classification task. Their interaction method allows users to gain better insight into complex structures of neural networks.

High dimensional visualizations with 3D representations benefit from metaphors that make knowledge more accessible and understandable. An essential constituent in computer sciences is the use of metaphors. The assessment of the metaphor is related to the properties of the visualized field and approximation of the notions of the related field. Design of the components related to specific features is crucial in this approach. One of the extensively used metaphors in computer sciences is the city metaphor. For example, the EvoStreets technique (Steinbeck et al., 2019) uses a city metaphor that visualizes hierarchical relations as software streets. VR City (Vincur et al., 2017) consists of different layers to hold various entities using a layout algorithm. It includes connection layers for relationships, an authors' layer to show recent activities with waypoints, a city layer to represent classes, a code space layer to scan codes, and a UI space layer for possible actions. Oberhauser and Lecon (2017) provide space, terrestrial, custom metaphors for fly-through experience to encourage exploratory, analytical, and descriptive cognitive processes on code information. In IslandViz, Misiak et al. (2018) utilize an island metaphor to visualize the software architecture of a software system based on the Open Service Gateway Initiative (OSGI) in VR.

## 4.4 Scientific Visualization

The recognition of graphics as a distinct field actualized with the achievements of powerful computers and photorealistic renderings has made scientists available to use visualization for scientific studies (Figure 5). Visualizing scientific data is crucial for experts working on scientific domains and communicating with a general audience and students. Scientific visualizations were limited to two-dimensional representations. With the development of rendering techniques, it is now possible to visualize and interact with scientific data in a 3D virtual world. This allows users to explore and interact with real-like representations that improve the comprehension capabilities of students and enhance public engagement.

Scientific visualization scans can be very complex and hard to compute according to the dataset due to high-dimensional and abstract data. They might require exclusive visualizations instead of traditional computer ones. Different 2D sections implemented with conventional desktop and mouse may not be sufficient to construct 3D understanding, which also differs according to the user. Unlike a monocular system like traditional 2D screens, binocular systems in virtual reality display provide a true sense of depth perception and spatial relations. Thus, many industries have adopted these systems to test real-life scenarios as in training. Serving for the mining industry software module, the block cave mining system visualizer was developed to contribute to the block cave's management cycle and operation. Allowing a collaborative environment and converting the complex context of the mining system into a graphical representation improves the understanding of seismic data (Tibbett et al., 2015). Interactive 3D data visualizations are also used in High Energy Physics (HEP) experiments. ATLASrift project (Riccardo Maria et al., 2019) aims to create an immersive experience for the atlas detector and experiment site. Utilizing Unreal Engine, they offer three levels with different modes of interaction. Although game engines are widely used for visualization when there is a need for an external library, as in the theATLASrift project, external libraries can be challenging to integrate into the game engines.

### *4.4.1 Meteorology and Earth Sciences*

Helbig et al. (2014) used VR to visualize atmospheric cloud data that allows their heterogeneous data in different dimensions to be visualized with the relationships between variables that otherwise cannot be easily understood by only showing the numbers. Advancing over prior meteorological visualization systems, MeteoVis (Li et al., 2020) offers concurrent spatiotemporal meteorological data streams from multiple sources with a wide range of manipulation and exploration features. Challenges of geosciences are generally related to visual requirements such as size, shape, and structure. The shader interface provided by game engines is used to visualize the terrain within the geophysical context. This allows rendering higher resolution meshes in the field of view to give details and conserve the extended view with low-resolution meshes. Bonali et al. (2022) reconstructed selected geological environments using photogrammetry techniques. According to their extensive user tests, most of the students and academics agreed on the usefulness of VR technology. They also draw attention to the accessibility of data and experience by promoting immersive technologies.

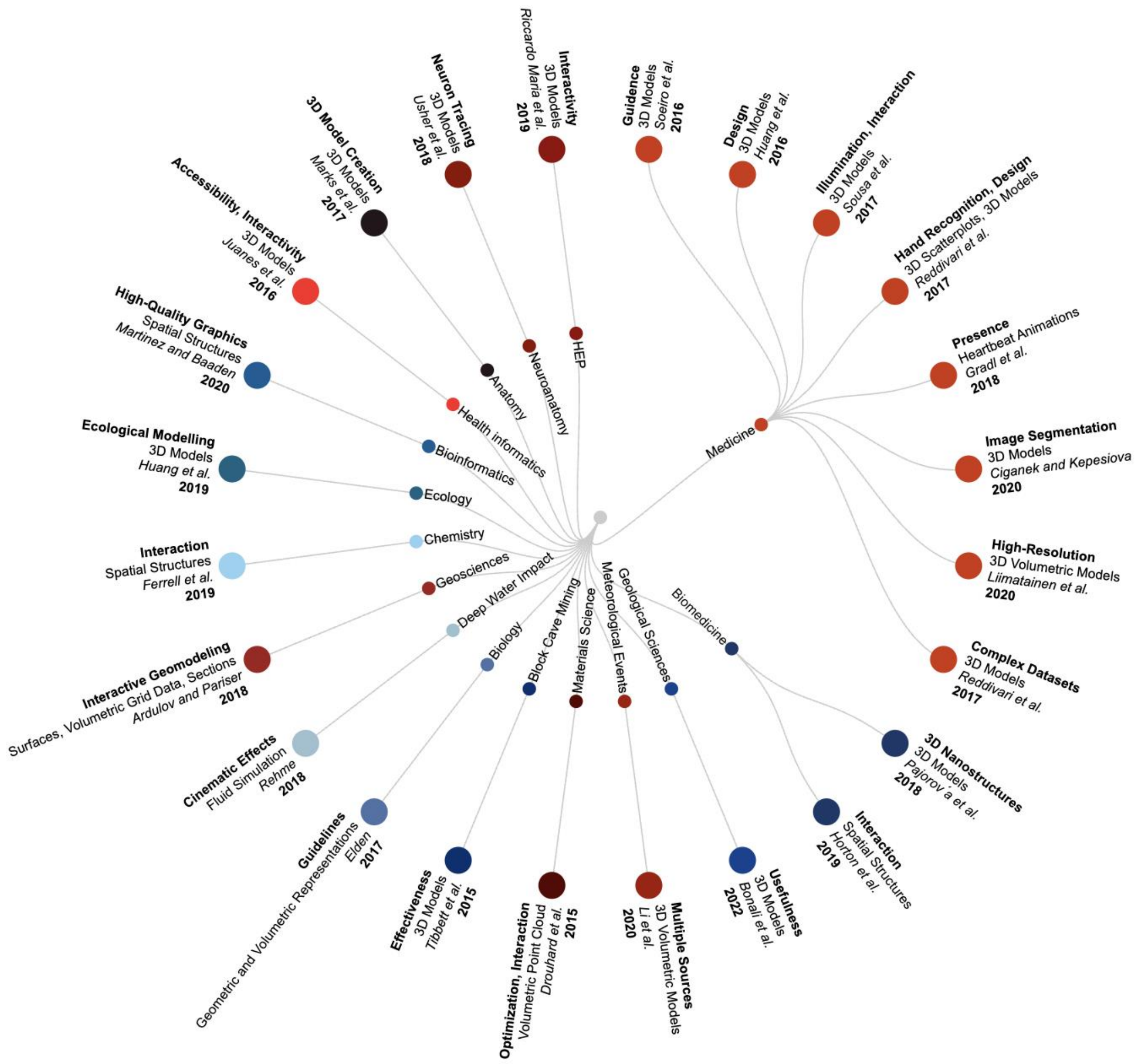

**Figure 5** Distribution of Scientific Visualization studies Produced with RawGraphs

Planetary-related models are inherently multilayered; therefore, maintaining a holistic approach for one-to-one 3D models may not be adequate. VRGE (Ardulov and Pariser, 2017) addresses this problem with surface viewing, volumetric grid data, cross-section viewing, and surface editing. Presenting various species under climate change scenarios, Huang et al. (2019) provide perspective views from different heights and filters to retrieve information. Subjectivity, artistic elements, use of ornaments have always been burning questions of the visualization sphere. Especially for scientific visualizations, the presentation must preserve the data's accuracy, integrity, and credibility. On the other hand, some examples used argumentative elements without spoiling the data's essence, contrarily enriching it by engaging. For instance, Rehme (2018) employed artistic and cinematic elements such as shaders, shadows, camera movements, and slow-motion, which bring new functionalities to the end product.

#### *4.4.2 Nanosciences*

With the development of advanced techniques, nanoscience has been proliferated. In the context of nanosciences, virtual reality has been employed for chemistry (Ferrell et al., 2019), materials sciences (Drouhard et al., 2015), biomedicine (Pajorov´a et al., 2018), bioinformatics (Martinez and Baaden, 2020; Sommer et al., 2018), solution finding for medicine, and health care (Gradl et al., 2018). Visualization and transfer of the structural properties of nanostructures require different techniques.

Being an advanced approach, scanning electron microscopy (SEM) is one of those techniques in which the obtained data can be transferred to virtual environments. There are several approaches to convert scientific knowledge to interactive 3D VR environment format. Conversion can consist of artistic representations, direct visualization of data using technologies like electron microscopy, or simplified 3D models. As in the case of geosciences, different artistic features provided by game engines are employed to amplify certain features of small-scale structures. GEARS (Horton et al., 2019) utilizes surface shader properties to emphasize selected features of confocal microscopy data. There are also different rendering techniques available for rendering. For instance, while ray-traced volumetric rendering composes the object via simulating the light, geometric rendering takes advantage of 2D sections to construct a 3D model.

Material sciences extensively rely on volume rendering. Therefore, to overcome latency, optimized rendering algorithms are employed. To create intuitive interaction and natural controls before transmitting volumetric point cloud representation to a game engine, Drouhard et al. (2015) offered the use of extraction methods to reduce the size and provide better optimization.

#### *4.4.3 Medicine and Biology*

Medical visualizations are generally composed of 2D mediums such as cross-sectional images and magnetic resonance imaging (MRI) scans. Therefore, 3D geometric models are reconstructed from individual slices (Reddivari et al., 2017; Soeiro et al., 2016; Juanes et al., 2016). Another method is image segmentation, which helps separate the pixels of the specified parts from the overall image. Ciganek and Kepesiova (2020) propose a segmentation and 3D model reconstruction method based on machine learning algorithms. Visualizations of medical or anatomical models consist of many complex subparts (Liimatainen et al., 2020). Thus, studies designed interfaces that include different features such as labeling, highlighting (Marks et al., 2017), distinct colors, selective visualization (Soeiro et al., 2016), and navigators (Juanes et al., 2016). Also, to solve depth perception problems in rendering photography effects, depth-of-field (DOF) (Martinez and Baaden, 2020) and gradient shading (Usher et al., 2017) are used to sustain depth cues. Virtual reality can assist the diagnosis process by providing appropriate conditions. To reduce the diagnostic errors in radiology, Sousa et al. (2017) designed a virtual reading room where the reader can adjust the illumination and ambient light while displaying luminance.

Some studies include different approaches and subjects such as medicine education(Huang et al., 2016), hand recognition (Reddivari et al., 2017), smartphone applications for anatomy (Juanes et al., 2016; Soeiro et al., 2016), and biofeedback (Gradl et al., 2018). Although game engines are employed by most of the studies due to graphics performance, physics, and ease of deployment to VR, game engines may not be proper for scientific visualizations, as examined in the study of Elden (2017). Elden (2017)'s study includes three demos; artery, rat brain, and genome. According to the requirements of the demos, different representation techniques were chosen. Thereby, the study became convenient to construct guidelines. As reported by the study, game engines are designed for geometric visualizations, and their priority is not accuracy but speed, which makes game engines unreliable.

### 4.5 Collaborative VR

Collaborative Virtual Environments (CVEs) provide remote and collaborative interaction on various data representations, independent from the users' physical location. Within a comprehensive spatial environment, they allow users to train, review and discuss using different information channels (Churchill and Snowdon, 1998). Previous CVEs included information visualization, teleconferencing, simulation, and social events. To design CVEs, the most commonly employed technologies were large spatial immersion displays and Virtual Environments (VEs), such as CAVEs and HMDs. These technologies have many differences, such as resolution, presence, and freedom of movement. Cordeil et al. (2017b) conducted a user study consisting of a series of tasks on 3D network visualizations to compare those in VR platforms. According to the results, while there is no major difference in accuracy and experience, HMDs offer faster interaction. Therefore, modern HMDs are preferable for immersive visualizations instead of expensive CAVE-style facilities. CVEs aim to provide enjoyment, social interaction, and presence, which provide cognitive benefits.

Collaborative environments aim to connect multiple users more naturally and increase users' awareness to break down the isolation. The environment can require participants to be co-located or remote and present different interaction levels, such as symmetric or asymmetric. Co-located studies are preferable due to network limitations. However, they enhance people's interactions within a single space and limit VR opportunities. Asymmetric applications do not offer the same interaction possibilities to all participants. For example, while one user interacts with using a VR head-mounted display, another user might experience the VR through traditional screens. The interdependence of users is inextricably linked to the special demands of visualization. Therefore, depending on the visualization and collaboration mechanics, different degrees of asymmetry can be used in the setup. The use of different devices has already produced an asymmetry in terms of visualization. According to the roles of the participants, this asymmetry arising from hardware differences can be a conscious design choice. For instance, ShareVR (Gugenheimer et al., 2017a) presents an experience where Non-HMD and HMD users can interact with each other and the environment together. They implemented several cases to construct guidelines for co-located asymmetric VR experiences. According to the results of their study, shared physical space and physical interaction enhance the experience and enable novel interaction methods for VR and VR games.

Another important interaction decision is the determination of viewpoint. This decision can be divided into two parts for collaborative environments. The first step is determining where a user is looking in the scene, and the second step is the decision of independence between users. Although the general tendency is to use the "what I see is what you see" principle with only one shared view, the environment can require or provide multiple views for different users either synchronously or asynchronously. For example, PlottyVR (Brunhart-Lupo et al., 2020) is composed of various statistical tools and libraries and offers a myriad of visualizations. Providing multiple viewpoints for different users enables each user to utilize different types of visualization. In some cases, the environment facilitates asymmetric collaboration and multiple viewpoints. Pointing out this requirement in design and architecture, Ibayashi et al. (2015) propose a system named Dollhouse VR that consists of a multi-touch tabletop device to manipulate the environment and HMDs to provide an internal view. Xia et al. (2018a) created a scene editing tool to support collaborative workflow. The tool, Spacetime, introduces three interaction concepts to enable users to easily manipulate the environment: container, parallel object, and avatar objects.

Companies that use the advantages of remote working have also implemented virtual reality to benefit from interaction possibilities that existing communication tools can not offer. Therefore, commercial virtual tools for teamwork have emerged, such as VISIONxR (Xia et al., 2018b). It allows multiple users in multiple locations, on multiple devices. These virtual platforms aim to improve the quality and effectiveness and create easily adaptable environment options and interfaces. Virtual collaboration is still less effective than sharing the same physical location in terms of expressive communication, including voice gaze, gestures, facial expressions, or full-body movements. According to Fussell and Setlock (2014), visual actions are way more important than speaking for communication in virtual environments. Therefore, to convey messages and provide more effective and efficient communication, avatars and hand gestures were started to be used in virtual environments to replace the expressions belonging to the physical world. To increase the sense of physical co-presence, Amores et al. (Amores et al., 2015) present an immersive mobile platform, ShowMe. Using depth sensors and cameras enables users to see their hand gestures and hands, making it easier to collaboratively work on a physical task.

VR devices need their hardware equipment; thus, collaborative environments are composed of not one but several worlds, each for one user. This situation creates two types of users; authorized users and connected users. While the authorized user has direct control over the world with local machines, the world of the connected user is continuously synchronized. This separation is an opportunity Hoppe et al. (2021) presented ShiSha, which uses shifted but shared perspective modification for the remote virtual environment. Therefore, it can enable users to observe from the same point of view while they can see other's virtual avatars in their virtual worlds. Having multiple users share a virtual space requires virtual representations of individuals. Another work that accentuates the potential of visualizations of avatars is Multi-User Cell Arena (MUCA), where (Bailey et al., 2019) offer customizable avatars. Embodiment is the foundation of many social VR experiences and positively affects presence. Customizable VR avatars have the potential to increase the sense of belonging.

### 4.6 Training and Simulation

VR provides the capability of training people from different professions to deal with complex situations and prepare them for their roles in real environments. It has become an essential training tool for soldiers, doctors, drivers, and pilots. Also, it is used for patient rehabilitation and disaster management. According to Ott and Freina ( 2015), the main motivation to use VR is its ability to provide experiences for context or environment that are inaccessible, problematic, or dangerous. Collection of reviews in the study of Mikropoulos and Natsis (Mikropoulos and Natsis, 2011) reported that in comparison with other systems, VR is more advantageous by an only sense of presence and dynamic 3D content, which have a positive impact on learning. Majority of the studies on training collected around the areas of medicine (Chang and Weiner, 2016), safety (Xu et al., 2017; Jeelani et al., 2020a), industry (Grabowski and Jankowski, 2015), and crisis and emergency management (Ronchi et al., 2016; Kwok et al., 2019; Molka-Danielsen et al., 2015; Surer et al., 2021), and rehabilitation (Joo et al., 2020; Yates et al., 2016).

As VR training scenarios mainly involve computer-generated 3D graphics, 3D modeling is important in creating virtual training environments. There are several options for developers to create related content. Advanced 3D modeling software, such as 3ds Max, Rhinoceros, Maya, and Blender, provide developers to create realistic environments. In addition to those tools, asset stores of the game engines and 3D model libraries present a variety of options. The reusability and customizability of 3D models decrease the cost of systems. To achieve high levels of immersion and presence essential for a training environment, artificial generation of actual information needs to stimulate the major senses. Therefore, to sustain a fully immersive experience, stereoscopic displays, motion tracking hardware, and input devices are employed. The release of motion-sensing controllers primarily developed for games such as Nintendo Wii Balance Board, Sony Playstation Move, and Microsoft Kinect has promoted the evolution of training. With the contribution of additional devices, VR training has expanded to a larger sphere, including dance training, aircraft controls, and rehabilitation. Investigation of virtual reality for practical use gave birth to new terms such as virtual factories, which are simulated models that consist of many sub-models to represent the cells of a factory. This integration offers planning, improvement of product, efficient planning phase, decision support, testing, and controlling the systems. In the context of Industry 4.0, different visualization techniques, dynamic virtual models, and types of simulation, such as discrete event or 3D motion simulation, are employed by automotive engineering, aerospace engineering, mechanical engineering, and medicine. VR training is often preferable for medicine since it offers emergency management, cost-effectiveness, recursiveness for tasks, and remote surgical training, which requires haptic devices due to physical procedures.

To manipulate virtual objects via haptic devices, in addition to geometry-based modeling, medical procedures require physics-based modeling to simulate deformable objects (Escobar-Castillejos et al., 2016). However, the animation of deformable objects in virtual reality environments is still a challenging problem, and an efficient physics-based method for virtual object interaction requires computational complexity. The physical interaction with the virtual objects needs to be realistically simulated to be convincing. This is especially important for the training scenarios that require detailed hand interaction, such as surgical training. Actual hand motion leads to unstable results in physics engines which cause interpenetration. Most proposed interaction approaches are simplified to avoid realistic simulations' complexity and error margin.

Recently, virtual reality has started to substitute traditional rehabilitation methods. Proposed methods for this transition blend the various hypotheses coming from different areas. For example, based on hypotheses stemming from neuromotor rehabilitation, game development and design took shape together in Rehabilitation Gaming System (RGS) and Personalized Training Module (PTM), which was developed to adjust task difficulty. According to intended results, rehabilitation requires specialized visualization techniques. For example, within the development of motor function awareness, vision is competent to give feedback, and also neuromotor rehabilitation may depend upon movement and environment visualization (Tsuji and Ogata, 2015). To meet these requirements, TRAVEE (Voinea et al., 2015) offers 3D scenes used for neuromotor rehabilitation and a user-friendly interface that positively affects users' processes.

Emergencies are unexpected events that require a rapid and effective response. To improve human behavior under artificial and natural disasters, simulations such as hurricane flood for analysis and control, fire safety, and earthquake simulations to assess human perception and behavior (Gamberini et al., 2015) have been designed and used. For a successful pre-evacuation or action, evaluation of the situation and the reaction time are key factors. Simulations of different emergency cases provide anticipation to assess the situation, awareness, and improvements in action time and behavior. For example, according to (Rosero, 2017), most participants present unsuccessful fire growth estimation results. The learning approach and nature of virtual environments made interaction an essential characteristic, including multifaceted features such as manipulation navigation. Additional to those features, personalization approaches and adaptive technologies should be preferred to increase the effectiveness of VR-based training (Jeelani et al., 2020b).

Despite the advantages of contributing to cognitive and psychomotor skills and helping users gain control over emotional response, virtual reality-based simulators that train individuals for high-risk industries remain questionable due to cybersickness and technological challenges. Areas such as aviation, fire-fighting, military, medicine, and manufacturing require a high level of realism to reach a certain level of success. Simulators may not efficiently represent uncertainties that result in oversimplified training environments. Another problem is that simulators are developed by software developers who are not experts on the selected subject most of the time. Therefore, the majority of studies in this branch focus on these problems. Vahdatikhaki et al. (2019) criticize most construction training simulators as unrealistic due to static site representations. Their framework offers four stages; context capture, context generation, context-user interaction, and context-based assessment. Although this process was proposed for construction site simulations, it is applicable for a wide range of cases that require context-realistic environments. Besides, the ease of use and reason behind the extensive use of game engines is achievements in approximation to reality, especially with the particles system tools offered. Examining fire simulations, it is certain that smoke's realistic spread and diffusion process is crucial. Smoke visualization requires high computer performance and a high level of realism, which can be provided through game engines.

Utilizing the advantages of game engines, Shamsuzzoha et al. (2019) propose a framework consisting of five stages from database to evaluation for industrial training and maintenance. Their prototype includes minimaps, blinking exclamation marks for attention, realistic visual effects, and an IoT screen to interact with the system, which are visualization preferences that make the interaction and information flow possible.

### 4.7 Web VR

Web services have become the primary data source, providing access to information anywhere and anytime. However, web browsers are limited in many cases, and most of the studies focus on solving those limitations. Due to the rendering capability of web browsers, presenting large-scale and real-time visualization demands a great deal of work. Yan et al. (2020) employed different online real-time fire training techniques to solve this problem. They prefer downloading the data gradually while the viewpoint changes and converting virtual people to lightweight versions. They employed a technique called "clone" rendering. In situ visualization or processing, the techniques where the data is visualized in real-time as simulation generates it, are used. Therefore, it does not involve storage resources; it is a natural solution for data transfer. Since it is a real-time generation, users can interfere with analyzing immediate effects.

VRSRAPID (Mascolino et al., 2019) web application uses extensible 3D virtual reality models (X3D) for interactive scientific computing. It is a collaborative and interactive environment designed for nuclear systems supported with real-time simulations. Traditional nuclear modeling and simulation tools are mostly built upon deterministic or statistical methods. Deterministic solutions are memory intensive and require significant computation resources. On the other hand, statistical approaches, such as the Monte Carlo method, may lead to statistical uncertainties.

To build an accurate and real-time model, they use RAPID Code System and generate the X3D models at the end of the calculation. Aiming to explore visualization methods for health data, Hadjar et al. (2018) propose a prototype application that utilizes several libraries and an A-Frame framework. Web analytics include charts, graphs, diagrams, animations integrated into visualization systems, and the combination of other visualization techniques. Using A-Frame allows developers to create interactivity based on ray-casting and animating the objects for efficiently interpreting a multidimensional dataset.

Libraries allow users to construct 3D visualizations by mapping the datasets that are obtained from external sources. Following the Shneiderman mantra, the web-based ExplorViz (Fittkau et al., 2015) tool presents a software city metaphor with gesture recognition for translation, rotation, zoom, selection, and reset tasks. Vria (Butcher et al., 2019) prefers 3D bar charts for data exploration and analysis due to their simplicity. With the increased importance of network environments based on VRGIS, the WebVRGIS engine (Lv et al., 2016) offers support for data publishing, transmission, and multiple users using and solving the problems of peer-to-peer (P2P) technology. Spatial analysis requires three-dimensional visualization of a largescale and multi-source urban landscape. Due to computation workload and required memory, rendering massive data in real-time is troublesome.

They used an interactive rendering system and visualization optimization technologies such as texture mapping, automatic level of detail, occlusion culling, and frustum culling to solve this problem. On top of this work, Li et al. (2016) offer to use the WebVRGIS engine to analyze and visualize real-time dynamic traffic data. As it is understood in the previous sections, most scientific visualizations rely on volumetric visualizations, especially in medical studies. To visualize mesh and volumetric data captured using 3D medical scanning in VR, Kokelj et al. (2018) developed a web-based application. Using the rendering pipeline of the Med3D framework, they used the volumetric ray casting technique, which performs calculations using output images.

NeuroCave (Keiriz et al., 2018) is a visual analytics tool that offers interactive methods and visualization choices for exploration. It enables users to distinguish regions and their functions using a color scheme. Instead of using realistic rendering methods, they construct a connectome using different platonic solids. To simplify the rendering process, ProteinVR (Cassidy et al., 2020) utilizes game-like camera movements where the objects are stable and only the camera can move. Thus, they were able to use pre-calculated shadows and textures to advance the performance in the browser.

## 4.8 Games, Visualization and VR

Video games are a collection of information and extensively rely on the presentation of information which holds various attributes that change according to state. Possessing a large amount of data, games played in the digital world are more complex and have different needs than games played in the physical world that are easily comprehensible. Data visualization has been used in games for generations to create continuous communication. While visualization components in older games are more straightforward and generally serve to transmit gameplay data, they are used for multiplexed situations in modern games. Representations like bar graphs and tree diagrams related to information visualization widely occur in games; however, especially the entertainment aspect creates a difference in implementation. Beyond the implementation details, the utilization of virtual reality in the video game industry has created the need for more radical differences. Although pioneering commercialized virtual reality video games presented with the release of Sega VR and Nintendo's Virtual Boy, they were considered unsuccessful in the nineties. The process began in 2016 and was followed by the introduction of various products such as Gear VR (Oculus), HTC Vive (HTC and Valve), PlayStation VR (Sony Interactive Media), and Samsung Gear VR, had achieved massive success by offering major novelties for video games.

Virtual reality games differ from traditional video games in terms of the level of immersion and type of interaction with virtual content. Decrease of connection-level between outer world and inclusion of body and hand movements provide innovative gaming experiences, which require new techniques for visualization of the virtual world. The interaction does not only occur between HMD users, but co-located participants can also interact. VR game ShareVR (Gugenheimer et al., 2017b) offers asymmetric interaction between HMD and Non-HMD users. Recent visualization researches and studies involve additional interface features, adaptive hints, context-sensitive tutorials, new approaches for player navigation. To make 3D manipulation easier in VR games, Rachevsky et al. (2018) offer to graphically represent the player's gestures as a part of the interface. According to (Polys and Bowman, 2004), instead of focusing on utility, visualizations in games should be functional and pleasing. Their developed framework proposes five elements to identify visualization techniques: primary purpose, target audience, temporal usage, visual complexity, and immersion.

### *4.8.1 Visual Realism and Presence*

Slater et al. (2009) proposed a division between components of realism as geometric and illumination. While geometric realism considers the properties of virtual objects, illumination realism deals with the convenience of lighting. Most of the existing studies and discussions recently conducted ground on Slater's theoretical framework of the place illusion (PI) and the plausibility illusion (PSI) in virtual reality and the sub-components of geometric and illuminations realism. Slater has argued on responses and defined two types of illusions based on the credibility of the events and the places. The PI mechanism in VR games endeavors to present game objects and places to increase presence. PSI mechanism offers persuasive game events and activities while the player actively engages with the simulated environment with a large field of vision. Beyond the visual, VR systems should offer auditory and haptic displays to sustain PI and PSI effectively. Impact levels of PI and PSI can vary according to the main objective of the environment. For example, Lynch and Martins' (2015) survey study examined the fright experience in immersive VR games. Later further categorized fear elements and identified the strategies and reactions towards fear elements. According to the results of their study, PSI elements trigger a higher level of fear response than PI elements. Another study conducted upon works of Slater complements (Hvass et al., 2018), the suggestions on effects of visual realism. Results of the physiological measures and self-reports of the participants revealed that a higher degree of geometric realism induces a stronger sensation of presence and emotional responses.

*4.8.2 Gameplay Data*

Gameplay-related data in the textual format is processed and presented through graphical representations that enable users to absorb the data. According to literature surveys (Sevastjanova et al., 2019; Wallner and Kriglstein, 2013) on gameplay data, charts and diagrams, heat maps, different types of movement visualizations, self-organizing maps (SOMs), and node-link representations are the most used types. Selection of the most efficient and convenient approach according to information to be represented is a crucial step. Although they can be interpreted in different forms, charts and diagrams are more suitable for direct demands than exploratory tasks.

According to the taxonomical study of Kriglstein (2019), there are two ways to collect gameplay data. Observation-based data can be collected by either observing the player interactions or using questionnaires and interviews, and it helps developers understand the players' motivations, behaviors, and preferences. Data can be collected through developed mediums automatically. One of the main differences between these approaches is that while the first one presents qualitative outputs, the second one produces quantitative data, which is more available for visualization. The data can be spatial, non-spatial, or temporal. The taxonomy of Kriglstein (2019) presents six different categories: comparison, distribution, relationships, time, space, and flow, based on tasks and types of data.

To our knowledge, unfortunately, there is not enough research to build a concrete understanding for building visualization for gameplay data in virtual reality yet. Visualization studies related to the presentation of the gameplay data mostly focus on traditional video games. Taxonomies and techniques built for information visualization are not available to adapt because of the unique needs of game data.

*4.8.3 Game Analytics*

The increasing complexity of games and audience called for new fields instead of traditional methods like user testing, play testing, surveys, and videotaping to evaluate player behavior. Designers, programmers, marketers, executives, and players are all using gameplay data in various ways and for different purposes. Using visualization is an inevitable option to digest data collectively. Although information visualization techniques are used, video games give various audiences a new direction to analytics with unique visual experiences. InfoVis community already has defined systems used for data analysis where analytics is not only focused, as "Casual Information Visualization" (Pousman et al. 2007)—belonging to this category, ambient, social, and artistic information visualizations are criticized for being unproductive. Visualizations integrate the play with data analysis considered as Playful InfoVis. Medler and Magerko (2011) have offered to broaden the scope and capabilities of Playful InfoVis.

Due to the nature of VR, analytics has become a vital component of VR games. Analytics helps developers increase performance by providing real-time information, fine-tuning via data presented, realizing problems in the design phase, and understanding player segments, players' engagement level, and playing style. Like gameplay data, game analytics are also not studied in the immersive visualization domain. Distinctively, game analytic visualizations produced for traditional environments can be used in immersive environments within the information visualization domain.

*4.8.4 Gamification and Gameful Concepts*

Even though it is not a new concept, "gamification" has always been considered a contentious term, and parallel terms are continued to be introduced in the game community. Deterding et al. (2011) proposed the definition of "gamification" as the use of game design elements in non-game contexts. Later in their survey, Seaborn and Fels (2015) define gamification as "the intentional use of game elements or a gameful experience of non-game tasks and contexts." Gamification uses game elements such as points, unlocking, achievements, leader boards, levels, virtual items, quests, avatars, collections, competition, or cooperation in non-game applications to strengthen user motivation. On the other hand, serious games are designed for additional non-entertainment purposes. Previous studies verify that visual properties affect the user's motivation in citizen science games (Curtis, 2015; Miller et al., 2019). EyeWire (Tinati et al., 2017) is a web-based gamified citizen science platform that encourages users to perform complex tasks by transforming them into more manageable tasks in a gamified environment.

According to a study (Tinati et al., 2016), gamification elements such as leaderboards, individual points, customizable roles, and visual appliances increase users' engagement. Foldit (Curtis, 2015) is another citizen science game that is a puzzle game that contains molecular visualizations on protein folding. Analysis of game-play data on view options settings displays significant differences in the visualization choices of experts and novices according to tasks (Miller et al., 2019).

With VR, AR, and MR technologies, data visualization transforms from passive to more interactive exploration. Combining the interactive nature of gameful design concepts with data visualization can reduce cognitive overload while immersing the players in the content. For this purpose, Wanick et al. (2019) provide two case studies on orbital visualization and earth data visualization for scientific data where they combine game design concepts and data worlds with VR technologies. To test the usability of VR game interactions in scientific domains, Bergmann et al. (2017) developed visualizations belonging to fields of particle physics, biology, and medical imaging appropriate for game interaction techniques in VR. They reported that although VR game strategies include some difficulties, such as simulation sickness, they offer a myriad of opportunities and a high level of immersion for scientific domains. GamefulVA (Sevastjanova et al., 2019) is proposed for fostering motivation by combining gameful design concepts with visual analytics.

Another genre that combines video games with other domains is exergames. Exergames aim to blend physical exercise and video games, requiring players to move physically due to gameplay mechanics. Beyond the non-immersive exergames played with controllers such as Wii Remote and Microsoft Kinect sensor, there are also pervasive games such as Ingress and Pokemon Go that also encourage physical activities. With the recent developments, VR has become more efficient in engagement and performance. Therefore, video games are converted into VR format, and new games that require body movement were released, such as Fruit Ninja VR, Hot Squat, Holopoint, and Portal Stories: VR. According to an evaluation study (Gugenheimer et al., 2017b), VR games reduce the perceived exertion, motivating people to exercise more. Additional to opportunities presented directly with VR, some studies focus on personalization and difficulty adjusting adaptable interfaces to keep the player engaged. Different strategies have been proposed for the visualization of physical activity. In an abstract information display named HappyFit (Yoo et al., 2017), they prefer to keep visualizations abstract, nonintrusive, positive via colors, shapes, metaphors to give consistent information and keep the players engaged. The primary aim behind this strategy is to encourage physical activity in the short and long term by increasing awareness, which creates a personal response.

### 4.9 Design Considerations and User Interactions

This section of the article aims to explain the importance of the evaluation methods and results with advice for design considerations in previous studies. The visualization field tends to rely on assumptions that are not proven but are firmly rooted. Kosara (2016) defines this situation as an "empire built on sand" and explains with his studies that even assumptions on most commonly used visualizations can be wrong. Questioning and testing assumptions repetitively to reach evidence is crucial to create a solid foundation for development and evolution (Kosara, 2016). As collective responsibility of the community, in the literature, hundreds of authors can be found who used and developed various metrics, taxonomies, and typologies, proposed guidelines, and created models for the multi-level understanding process behind and making assessments for different aspects of visualizations.

#### *4.9.1 Visual Perception*

Although perception is extensively studied in VR, new perceptual challenges continue to proliferate. Those challenges not only include the 3D visualization features such as depth and distances, shapes, sizes, colors, and contrasts but also involve hardware-related problems. 3D human visual perception is closely related to depth perception, which also can define the effectiveness and comfort level of 3D visualizations (Dede, 2009). Visual cues provide depth perception in 3D environments, such as occlusion, rotation, shadows, and shading. Viewing 3D visualization through multiple angles enhanced the understanding of the data. Unfortunately, rotational aspects are not suitable for data presented in textual format. According to Bertin (1983), three levels of human visual perception can be described. Individual elements, groups of elements, or whole images can be the subject of focus. The ability of focus sustains users to complete specific tasks. According to the density of the data, raw data should be converted in a way that helps users' focusing abilities. To achieve this, methods can include visual attributes like color, position, size, shape, and techniques, such as different perspectives or clustering.

Visual perception in virtual environments generally has been studied with the lenses of Gibson's theory. Gibson (1977) stated that the environment offers different action possibilities to actors, called affordance. According to this idea, actor and environment coexist, and perception is directly related to action. From this perspective, researchers have studied perceptions of affordance for different situations, height, and depth perception. Cliquet et al. (2017) analyzed the perception of affordances while standing on a slanted surface in VR, and also they considered the effects of different materials. According to the results of their experiments in VR, although participants could discriminate the angles appropriate for upright positions, the approximate critical angle for an upright posture determined by participants is lower than the results of the studies conducted in real environments. Nagao et al. (2018) designed an interface for infinite stair demonstration with passive haptic slats and markers to increase a sense of presence and riser height. Later, to examine height perception while moving, Asjad et al. (2018) designed an infinite ascending staircase for the virtual environment, which shares exact dimensions with a staircase in the physical world. According to their study, virtual shoes positively affect presence and error estimation.

#### *4.9.2 Movement*

Sustaining efficient navigation for users in the virtual environment has been a challenge. Although the most natural virtual locomotion technique is mapping users' physical movements directly, different locomotion techniques have been offered due to the limited physical space.

Teleportation is a locomotion technique that generally requires the user to aim for the target location in a virtual environment. Although it overcomes spatial constraints and provides users to travel in multi-user experiences, it may cause confusion and lack of constant feedback due to spatial discontinuity of users due to teleportation. To solve this problem and maintain communication between users, Thanyadit et al. (2020), considering time efficiency, traceability, intuitiveness, and recognizability, propose four different visualizations: hover, jump, fade, and portal. Those representations of movements aim to create traceable visualization to give feedback to other users to avoid confusion. The cues such as traces are helpful for users to understand their location and decrease the spatial cognitive cost. Cherep et al. (2022) conducted a study to understand the effects of teleportation interfaces on different individuals. Their results suggest that the design of the interface and individual differences create diverse spatial cognition cases, and concerning those parameters, users' awareness of location changes. This study points out the importance of defining the target audience and knowing the effects of design choices in immersive environments.

#### *4.9.3 Interaction*

Immersive technologies have introduced many new challenges to researchers. Interaction techniques can make the experience more effortless or cumbersome. Meaningful interaction between the audience and the visualized data enhances the immersion in a virtual world where the audience can see the data and explore different aspects. Research has produced a myriad of interaction techniques for VR, such as selection, manipulation, and locomotion. Researchers are searching for creative ways for visualization tasks to interact with data.

Onorati et al. (2018) developed an immersive bubble chart, especially to get information from unstructured data. Their work includes category bubbles that semantically group individual bubbles and allow users to explore data through words. An experience designed in Unity allows the user to grab, zoom, remove and merge bubbles and keep track of previous views. A First-Person Shooter (FPS) game for both HMD with specialized controls and non-immersive games that utilize monitors and have traditional controls was developed and evaluated by Rachevsky et al. (2018), and the different versions of the game were tested. While the free aim version has better results than the immersive one, since the users moved the camera with the keyboard in the free aim version, the fixed aim version has better results for the non-immersive versions. In terms of usability, results are different for immersive and non-immersive cases, and they agree that more natural and intuitive interactions are necessary for immersive games.

### 4.10 Comparative Studies

Design and implementation decisions have significant importance for data presentation. The process starts with the raw data, and until it takes the final form, many choices have to be made. The emergence of new technologies, techniques, and ideas requires continuous analyses to create a concrete background. Unfortunately, immersive visualizations suffer from the lack of standard guidelines to build upon. To prove the rationale behind the preferences, comparative studies were examined. Immersive technologies imply specific outcomes when they are tested for user experience. There is a threshold requirement to pass to choose immersive visualizations over traditional ones. This threshold is closely related to widely studied questions in HCI, which are user experience and technology acceptance. For example, Shrestha et al. (2016) recreated historical sites in Nepal for the CAVE environment to compare with the Paper-Based Artifacts. Results indicated that participants had difficulties solving complex problems in VR. This struggle can be the reason and result from the resistance originating from not being accustomed to new technologies.

Ren and Hornecker (2021) conducted a user study to compare virtual and physical data representations by creating two equivalent representations of the same data set for physical and virtual environments. This study showed that while physicalization helps decrease response time, participants tend to move slowly in the VR environment due to VR lag, which affects the quality of experience. According to a study conducted by Millais et al. (2018) on scatterplots, the workloads of traditional visualizations and visualizations in VR are almost equal. They reported that users feel more satisfied and successful when using VR for data exploration. Task-based comparisons of the 2D and 3D versions of visualizations give more diverse results. According to results of a comparative study (Kraus et al., 2020), which utilizes visualization tasks taxonomy by Brehmer and Munzner for the overview tasks, 2D heat maps gave better results while reading and comparing single data items, and 3D heat maps tested in virtual reality environment showed lower error rates. This situation leads researchers to use hybrid techniques.

In their study, Roberts et al. (2022) discuss different visualization techniques using case studies to identify the key features to create diverse 3D visualizations. Their results show that using multiple views and different viewpoints enhanced the understanding. However, each visualization technique requires solving a variety of problems. Those problems are not always related to the structure of the visualization. Immersive visualizations require optimizing several parameters at the same time. Beyond those comparisons, finding the effective use of space to represent data in immersive environments is another research subject. A Meta-analysis study (Akpan and Shanker, 2019) consists of 162 synthesized studies. VR offers more effective model development, verification, and validation performance on DES task performance when compared with the 2D visualizations. There are several approaches to guide users in virtual environments.

Guidance can occur via sounds, visual cues, or animations. For example, annotations in virtual environments are generally represented in abstract gestures, texts, or simple objects for indication. Alternative guidance can be based on imitation using a virtual tutor, which demonstrates the task that needs to be done. Based on three different tasks, Lee et al. (2019) compared the effectiveness of Annotation and Tutor. According to overall results, Annotation was found more helpful for accuracy and time performance. On the other hand, using a tutor improves recalling the pattern.

## 5 Discussion

Due to vast options and lack of common ground, selecting or creating methods and techniques to visualize a particular type of information is challenging. Although the main aim of all visualization techniques is communication via visual medium according to data, user, defined tasks, and presentation techniques starting from the data gathering, the whole process requires different techniques. Most scientific visualization data are obtained through devices such as sensors, microscopes, cameras, or via simulations volume rendering, slicing to extract sections. Therefore, data is already suitable to be transferred to virtual 3D environments.

Data visualization studies are more related to abstract data and focus on presentation and analysis aspects. Considering studies, according to the subject, information visualization can include aspects of both. For example, techniques used in art and architecture domains have more similarities with scientific visualization techniques, while computer sciences deal with similar problems with data visualization. Specific requirements of the domains create different discussions on various subjects. For example, nano-sciences and architectural heritage domains already have 3D physical forms to transform. For this transformation, discussion topics can be clustered around the level of abstraction, interaction techniques for the model, or presentation of textual data. On the other hand, if statistical data on cultural heritage is subjected to visualization, this time, data does not have a physical equivalent to display it visually. Physical attributes will be attended to features of data and design principles; color and perception theories may have a more dominant role.

Physical attributes and perception theories are widely used for abstraction. Considering the interaction possibilities presented by data itself, generally, there is a correlation between abstraction level and presented interaction capabilities. Abstract visualizations are more open to manipulation and interaction and more suitable for analysis.

On the other hand, more literal and realistic visualizations are more representative. Scientific visualizations mostly rely on 3D volumetric data combined with abstract elements. Although domain and subject are different according to data type, common visualization structures can be used. The combination of dynamic visualizations with time-series data is widely used in both information and scientific visualizations. 3D versions of bar graphs, line and scatter plots are still the most commonly used visualizations in various domains. The common objective is finding the best user-friendly navigation and interaction techniques in an immersive environment. Therefore, studies belonging to different research areas have common problems and challenges considering visualization. In general, most of the research questions gathered around concepts of representation, perception, interaction, locomotion, and decision-making.

In the information visualizations community, cultural heritage and construction popularity is increasing due to CAD-based and BIM-based models. Therefore, interaction with the integrated models gains importance. In addition to interaction with 3D models and multi-dimensional visualization, collaborative studies present various opportunities to users. Most real-world cases require possessing sophisticated and expensive equipment or being in a dangerous environment. Scientific visualizations, simulations, and visualizations built for training serve as a replacement for real-world equipment and situations. Educational visualizations are also expanding their scope, and including game elements, they create more active and dynamic relations. Gamification motivates learners, and a combination of visual senses together with physical interaction improves users' recall mechanism. Movements, embodiment, and gestures also increase the level of understanding and provide more active roles. The learners have the opportunity to interact with objects of different scales. VR breaks space limitation: therefore, the chosen subject can include objects of any size. The scale of the data representations can be arbitrary and controlled by the user to interact with the representation at different levels. While users can prefer large-scale visualizations that surround them or manipulate them more quickly for a specific task, data can be room-scale or smaller. By changing the scale, users move between egocentric and exocentric approaches, suitable for different tasks. Egocentric visualizations provide more immersive experiences since data surround the user, fly through it, and demand less cognitive load. On the other, exocentric visualization gives better results in analytical tasks.

As the interaction methods are developed, chronological observation of visualization taxonomies shows that the scope, the number of tasks, and primary objectives also continuously expand. A correlation can be constructed between this expansion and new data extraction techniques and technology development. For example, while advanced techniques provide new data types, technologies like haptic devices, sensors, AR, VR, MR define new interaction techniques. There is no comprehensive visualization taxonomy study specialized in VR to our knowledge. At the same time, even though previous taxonomies have certain limitations due to similarities, they can be used as a base to construct taxonomies for VR.

Virtual reality technologies (Figure 6) increase the number of interaction techniques and transform the existing ones. For example, hands-free pointing either by eye gaze or with a head pose is an interaction method that comes into literature with VR—presenting the third axis rotary motion together with physical immersion. Users can perform locomotion by walking in the physical environment, using controllers or teleportation methods. VR comes with additional aspects that need analysis and classification to broaden the existing taxonomies. The visualization community needs more comparative studies showing the clear distinctions between traditional visualization techniques and VR visualizations to create objective rules and a baseline for design decisions.

Classifying and defining the target users are crucial for taxonomies, guidelines, and classifications to be produced. According to cognitive capabilities, technology use, and special situations of the users, decisions should be taken considering interaction and visualization preferences. There is a need for comparative studies to identify common patterns and specific groups to construct such knowledge. As can be seen in the related section, collaborative studies can produce controversial results. Especially, different results in the studies that compare VR visualizations with desktop environments stand out. Even the effect of different visualization mediums on spatial understanding is still debatable. This situation may stem from the differences in participants' or design-specific problems. Therefore, until the results converge at the same point, they should be repeated. The user-centered design process is essential for visualization tasks.

After determining guidelines, tools can be produced to create more accurate visualizations. This way, visualization studies can gain speed and find a common language and visual consistency. Having previously tested and proved de facto standards is also helpful to accustom users to new technologies and switch between them. Unfortunately, developed tools are limited to specific data and visualization types. For example, most toolkits are produced to represent quantitative data, and their scope is limited to quantitative analysis. Also, it is a more complicated task to create toolkits for the qualitative analysis. Tools need to be flexible to create possible solutions and different interaction methods since design preferences also direct the user behavior.

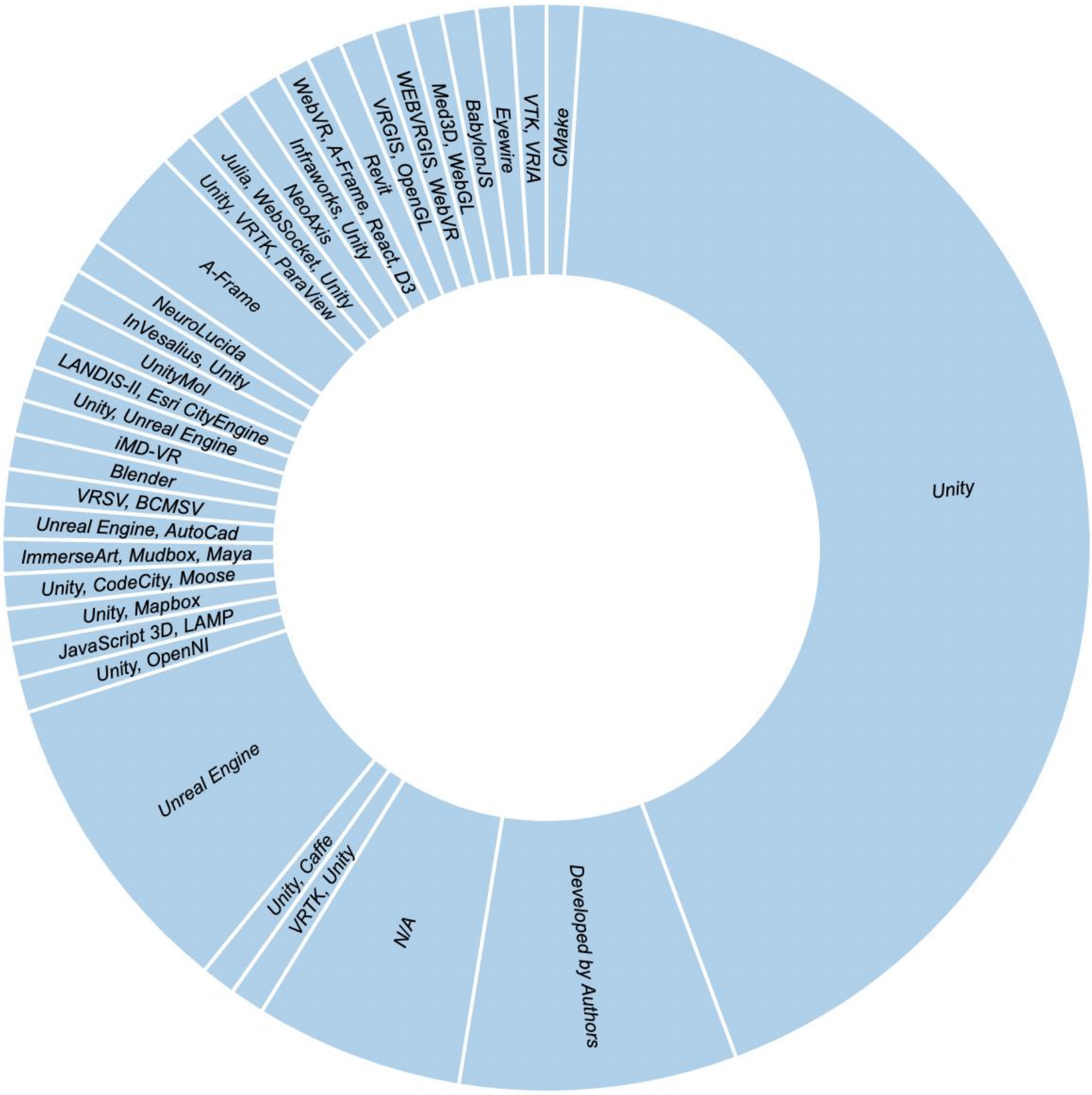

**Figure 6** The proportion of VR visualization tools among the selected studies

Virtual reality still suffers from technology acceptance and, due to hardware requirements, it is still not common. WebVR, VR games, and collaborative studies are essential in breaking the acceptance barrier. Recent trends are promising for the increase in virtual reality systems. Especially recent orientation towards remote collaboration paved the way for VR versions. Other examples can create a high degree of immersion, such as free-viewpoint video technologies that can be adapted for both WebVR and smartphone visualizations. Smartphones are being widely used in almost all aspects of our daily activities. Also, creating mobile versions of visualizations resolves the need for hardware. Simpler hardware such as Google Daydream and Samsung Gear VR offer controllers for mobile VR. However, mobile VR is not suitable for complex visualizations due to technological requirements since the performance is entirely based on smartphones. Unfortunately, applications using the web browser also have rendering and speed limitations. Even so, for specific areas, they can be beneficial.

Exergames take the lead in visualization studies in the game domain. Most studies encourage users to exercise regularly and engage with the activity. Since they target users with specific needs, visualization and interface requirements are variable and require elaboration. To reach further potentials of VR integration in daily life activities, as in the exergames, can overcome the usage barriers.

Another subfield open to further developments to overcome acceptance problems is collaboration studies. They can include a combination of synchronous or asynchronous, on-site or remote, and symmetric or asymmetric approaches. Asynchronous and remote collaboration is crucial for scientists working together on the same project. Synchronous and on-site ones are primarily used in training and architectural studies. Asynchronous and on-site version is the less common version, but their scope can include architectural projects. For example, to create more efficient smart cities, users can engage through visualizations in specific hubs in the city. The combination of synchronous and remote allows collaborators to work together in real-time while they are in different locations. The synchronous aspect of the collaboration helps users not to feel isolated. Therefore, the acceptance problem is one of the essential features to overcome. Multi-user experiences have more positive effects on users compared to single-user versions. It increases the users' engagement and success rate. The combination of Web, collaboration, and mobile technologies can provide quick and easy dissemination of information.

Although most of the studies provide solutions for particular problems of VR, problems relating to virtual reality hardware have not been solved completely. Most of the time, problems of cybersickness, latency in tracking, low refresh rates disturb the users and break presence. Studies try to fixate the experience on a fast frame rate to avoid VR sickness. While this approach works for simple visualizations, they follow different complex visualizations and simulations techniques. Allowing users to transport physically in a consistent environment allows VR to create complete immersion that makes it distinctive. However, the design of the visualization can also create fluctuations in the user's sense of presence and break the immersion. Inside VR, the user is only able to see computer-generated imagery. Therefore, the objects the user stares at should be more apparent than objects farther away or nearer.

Visualizations can also use focusing techniques to create hierarchy levels to differentiate essential aspects of data. Geometric realism affects the presence in VR and can create a strong presence and emotional response. Polygon count and texture resolution are generally increased to create more realistic environments. However, intended realism levels require powerful hardware. Therefore, different techniques are used, such as foveate rendering and occlusion culling. In immersive environments, the user can see all directions; therefore, different from the 2D representations, 3D mapping techniques can be applied to data. Visibility of the presentation is also vital for the user. When the data includes complex relations, developers can use different rendering or layering techniques. For example, edges and nodes can be dense if a network visualization represents the data. To lower the occlusion level, developers can prefer to play with scale, line width, or divide the data into more layers. The crucial point is conveying the data without deterioration. To ensure correct and practical interpretation of data, proper and explicit visual encoding of the information is crucial.

Scale is another aspect that should be taken into consideration for precision. For example, large-scale maps are designed for estimation or extraction value tasks; some parts can be diminished according to visual encoding. Perceived objects sizes are also changed according to avatar realism and scale. The effectiveness of a graph representation of a dataset can be objectively evaluated according to visual features. The features of association and selection enable people to perceive and discriminate against particular objects or features. Visual features should be consistent throughout the whole visualization experience. Graphical perception studies that measure users' performances across visualization designs and different displays offer other studies to new insight into the utility of depth, color, geometries, and scale.

To encode spatial data according to positional aspects, data elements can be placed in the virtual coordinate system or placed on constructed maps or globe view, according to data. In parallel to this selection, developers select different interaction methods. Globe visualizations include rotational movement. However, users can only see a particular part of the globe. Therefore, while information visualization prefers to use placement on maps, scientific visualizations employ exact shapes or coordinate systems due to the accuracy aspect. Different 3D shapes, colors, opacity levels, and textures are used to represent located elements. There are also data-specific preferences, such as a combination of rotational models with a network diagram. Spatio-temporal visualizations are mainly used for simulations. One of the most used visualization techniques for multi-dimensional data is 3D scatterplots. Most of the immersive tools and toolkits provide immersive scatterplots construction. Since they are already available for various visual channels, the encoding process does not require additional techniques. However, it is not suitable for complex datasets for specific tasks due to visual clutter. To solve this problem, filtering methods are used. For selection tasks, ray casting is still the most used method. These methods allow a user to select a single object, draw a bounding box, or brush multiple objects that can be selectable. Interactive visualizations always need to give necessary feedback and visual cues to point out the possible interactions. For example, selected objects should be visually identifiable.

Data quality, handling with data streams, and semantic relation extraction from raw data remain a challenge. Although machine learning algorithms are promising, automatization of the process may yield inaccurate visualizations. Providing compact representations for large volumes of multivariate data is still waiting for more advanced techniques. Although studies generally employ supervised and unsupervised machine learning algorithms and dimensionality reduction techniques, they also report that the results can be misinterpreted or poorly handled. Thus, efficient and effective methods for compression and feature extraction are needed. For scientific visualizations, accuracy gains more importance to filter data or calculate probability. Generally, fine-tuning the hyper-parameters is used to arrange the sensitivity of the algorithm.

## 6 Conclusion

Following the protocols for a systematic review, sources related to immersive visualizations were identified and filtered out. The resulting articles were further analyzed, and relevant studies were grouped according to the most common problem domains and represented in different sections. We presented the overview of existing literature, discussed the strengths and weaknesses of the described methods, and pointed out unsolved problems and challenges. The results of our study show that there is a growing body of research examining immersive visualizations in a wide variety of problem domains. However, while most of the studies have been gathering around the development of immersive visualizations, only a minority have focused on building theoretical background. Most studies developed visualizations based on data type and domain and later tested for the visualization tasks. However, those tasks also have certain specifications. Using tasks only to evaluate the visualizations eliminates possible integration options and task-specific design decisions. Studies use similar interaction methods provided by controllers. This situation is not sufficient to meet the requirements of specific areas. New interaction paradigms that exploit full-body interactions should be searched.

Within the realm of visualization, the most mature areas of research have come from training, architecture, and game technologies. One of the most striking results is that although game engines are widely used in various domains, only a few studies target gameplay data visualizations for VR games. Many studies focused on how VR could be used for training and the requirements for increasing effectiveness. Already having a 3D presentation, most of the architectural studies have accomplished the transition process. Especially, the use of immersive systems to maintain cultural heritage in digital mediums is increasing. Most of the visualization studies utilize games technologies and novel techniques derived from computer graphics and human-computer interaction. With these technologies, the visualization domain has given rise to new research questions, followed by usability, interactivity, and reliability issues.

2D visualizations still protect their ground in displaying statistical and abstract data in information visualizations, while 3D visualizations are commonly used in physical sciences, engineering, and design. 3D versions of traditional statistical visualization techniques, such as bar plots and scatter plots, are still commonly used in data visualization contexts. However, only a few studies focus on creating standard guidelines for virtual reality, and each study individually provides a framework or employs previous studies on traditional 2D visualizations.

With the myriad of advantages provided for visualization and virtual reality, most studies prefer to use game engines. However, accuracy requirements are not convenient for critical scientific studies. Due to occlusion problems, perceptual distortion, absence of a common baseline, and noneffective 3D representations of abstract data on standard 2D monitors, validation of 3D is still waiting for further research and alternative approaches for solving design challenges. As in the life sciences, a more holistic approach is their hybrid use, which can compensate for each other. Virtual Reality visualizations have made some compelling advancements in recent years, but there is still plenty of room for improvement and further exploration. Many research questions are waiting for comprehensive studies: Which data types and visual structures are more suitable for VR, and are they increasing the performance of qualitative and quantitative data analysis? Is there a need for analytical methods specific to VR, and how can they be produced?

The constant changes in the underlying technology require repeated studies and constant development in theories behind the empirical studies. Together with the promising nature of VR, the quality and sophistication of visualizations are waiting for further improvements.